%% file: ReducedInclusions-arXiv.tex
\newif\ifpreprint
\preprinttrue
\ifpreprint
	\documentclass[onecolumn,draftcls]{IEEEtran}
	\usepackage[short,bysection]{Theorems}
	\usepackage{cite}
	
\else
	\documentclass{cocv}
\fi
\usepackage{amsmath, amssymb, amsfonts, mathtools, xargs, tensor, units, stmaryrd, mathrsfs, enumitem, xcolor, comment}
\usepackage[hidelinks]{hyperref}
\input{Macros.tex}

\ifpreprint
	\includecomment{preprint}
	\excludecomment{publishedArticle}
\else
	\excludecomment{preprint}
	\includecomment{publishedArticle}
\fi

\begin{document}
	
\title{On reduction of differential inclusions and Lyapunov stability}

\begin{publishedArticle}
	\thanks{This research is supported in part by NSF award numbers 1509516 and 1508757, ONR award number N00014-13-1-0151, AFRL award number FA8651-19-2-0009, and AFOSR award number FA9550-15-1-0155. Any opinions, findings and conclusions or recommendations expressed in this material are those of the authors and do not necessarily reflect the views of the sponsoring agency.}
	\author{Rushikesh Kamalapurkar}
	\address{School of Mechanical and Aerospace Engineering, Oklahoma State University, Stillwater, OK, USA;\\e-mail: rushikesh.kamalapurkar@okstate.edu}
	
	\author{Warren E. Dixon}
	\address{Department of Mechanical and Aerospace Engineering, University of Florida, Gainesville, FL, USA.}
	
	\author{Andrew R. Teel}
	\address{Department of Electrical and Computer Engineering, University of California, Santa Barbara, CA, USA.}
	
	
	
	\begin{abstract}
		In this paper, locally Lipschitz, regular functions are utilized to identify and remove infeasible directions from set-valued maps that define differential inclusions. The resulting reduced set-valued map is point-wise smaller (in the sense of set containment) than the original set-valued map. The corresponding reduced differential inclusion, defined by the reduced set-valued map, is utilized to develop a generalized notion of a derivative for locally Lipschitz candidate Lyapunov functions in the direction(s) of a set-valued map. The developed generalized derivative yields less conservative statements of Lyapunov stability theorems, invariance theorems, invariance-like results, and Matrosov theorems for differential inclusions. Included illustrative examples demonstrate the utility of the developed theory.
	\end{abstract}
\end{publishedArticle}

\begin{preprint}
	\author{Rushikesh Kamalapurkar, Warren E. Dixon, and Andrew R. Teel\thanks{Rushikesh Kamalapurkar is with the School of Mechanical and Aerospace Engineering, Oklahoma State University, Stillwater, OK, USA. {\tt\small rushikesh.kamalapurkar@okstate.edu}.}\thanks{Warren E. Dixon is with the Department of Mechanical and Aerospace Engineering, University of Florida, Gainesville, FL, USA. {\tt\small wdixon@ufl.edu}.}\thanks{Andrew R. Teel is with the Department of Electrical and Computer Engineering, University of California, Santa Barbara, CA, USA. {\tt\small teel@ece.ucsb.edu}.}}
\end{preprint}

\maketitle

\section{Introduction}
Differential inclusions  can be used to model and analyze a large variety of practical systems. For example, systems that utilize discontinuous control architectures such as sliding mode control, multiple model and sparse neural network adaptive control, finite state machines, gain scheduling control, etc., are analyzed using the theory of differential inclusions. Differential inclusions are also used to analyze robustness to bounded perturbations and modeling errors, to model physical phenomena such as coulomb friction and impact, and to model differential games \cite{SCC.Filippov1988,SCC.Krasovskii.Subbotin1988}.

Asymptotic properties of trajectories of differential inclusions are typically analyzed using Lyapunov-like comparison functions. Several generalized notions of the directional derivative are utilized to characterize the change in the value of a candidate Lyapunov function along the trajectories of a differential inclusion. Early results on stability of differential inclusions that utilize nonsmooth candidate Lyapunov functions are based on Dini directional derivatives \cite{SCC.Roxin1965,SCC.Paden.Sastry1987} and contingent derivatives \cite[Chapter 6]{SCC.Aubin.Cellina1984}. For locally Lipschitz, regular candidate Lyapunov functions, stability results based on Clarke's notion of generalized directional derivatives have been developed in results such as \cite{SCC.Shevitz.Paden1994,SCC.Bacciotti.Ceragioli1999,SCC.Hui.Haddad.ea2009}. In \cite{SCC.Shevitz.Paden1994}, Shevitz and Paden utilize the Clarke gradient to develop a set-valued generalized derivative along with several Lyapunov-based stability theorems. In \cite{SCC.Bacciotti.Ceragioli1999}, Bacciotti and Ceragioli introduce another set-valued generalized derivative that results in sets that are pointwise smaller than those generated by the set-valued derivative in \cite{SCC.Shevitz.Paden1994}; hence, the Lyapunov theorems in \cite{SCC.Bacciotti.Ceragioli1999} are generally less conservative than their counterparts in \cite{SCC.Shevitz.Paden1994}. The Lyapunov theorems developed by Bacciotti and Ceragioli have also been shown to be less conservative than those based on Dini and contingent derivatives, provided locally Lipschitz, regular candidate Lyapunov functions are employed (cf. \cite[Prop. 7]{SCC.Ceragioli1999}).

In this paper, and in the preliminary work in  \cite{SCC.Kamalapurkar.Dixon.ea2017}, locally Lipschitz, regular functions are utilized to identify and remove infeasible directions from a set-valued map that defines a differential inclusion to yield a pointwise smaller (in the sense of set containment) set-valued map that defines an equivalent reduced differential inclusion. Using the reduced differential inclusion, a novel generalization of the set-valued derivatives in \cite{SCC.Shevitz.Paden1994} and \cite{SCC.Bacciotti.Ceragioli1999} is introduced for locally Lipschitz candidate Lyapunov functions. The developed technique yields less conservative statements of Lyapunov stability results (cf. \cite{SCC.Roxin1965,SCC.Paden.Sastry1987,SCC.Shevitz.Paden1994,SCC.Michel.Wang1995,SCC.Bacciotti.Ceragioli1999,SCC.Moulay.Perruquetti2005}), invariance results (cf. \cite{SCC.Ryan1998,SCC.Logemann.Ryan2004,SCC.Bacciotti.Mazzi2005,SCC.Hui.Haddad.ea2009}), invariance-like results (cf. \cite[Thm. 2.5]{SCC.Haddad.Chellaboina.ea2006},\cite{SCC.Fischer.Kamalapurkar.ea2013}), and Matrosov results (cf. \cite{SCC.Matrosov1962,SCC.Loria.Panteley.ea2005,SCC.Sanfelice.Teel2009,SCC.Paden.Panja1988,SCC.Teel.Nesic.ea2016}) for differential inclusions.

The paper is organized as follows. Section \ref{sec:notation} introduces the notation. Sections \ref{sec:diff inclu} and \ref{sec:Set-valued-derivatives} review differential inclusions and Clarke-gradient-based set-valued derivatives from \cite{SCC.Shevitz.Paden1994} and \cite{SCC.Bacciotti.Ceragioli1999}, respectively. In Section \ref{sec:Reduced-Inclusions}, locally Lipschitz, regular functions are used to identify the infeasible directions in a set-valued map that defines a differential inclusion. Section \ref{sec:generalized-time-derivatives} develops a novel generalization of the notion of a derivative in the direction(s) of a set-valued map. 
\begin{publishedArticle}
	Section \ref{sec:Stability-of-Nonautonomous} states stability theorems, invariance-like results, and Matrosov theorems for differential inclusions using the developed novel definition of a generalized derivative.\footnote{An extension of the framework developed in this paper that generalizes LaSalle's invariance principle for time-invariant differential inclusions is available in \cite{arXivSCC.Kamalapurkar.Dixon.ea2018}.}
\end{publishedArticle}
\begin{preprint}
	Sections \ref{sec:Lyapunov-stability-theory} and \ref{sec:Stability-of-Nonautonomous} develop stability theory for autonomous and nonautonomous differential inclusions, respectively, using the new generalized derivative.
\end{preprint}
Illustrative examples where the developed stability theory is less conservative than results such as \cite{SCC.Shevitz.Paden1994} and \cite{SCC.Bacciotti.Ceragioli1999} are presented. Section \ref{sec:Conclusion} summarizes the article and includes concluding remarks.

\section{Notation}\label{sec:notation}

The $n-$dimensional Euclidean space is denoted by $\R^{n}$, $\mu$ denotes the Lebesgue measure on $\R^{n}$, $\mathcal{D}$ denotes an open and connected subset of $\R^{n}$, and $\Omega\coloneqq\mathcal{D}\times\ropen{0,\infty}$. Elements of $\R^{n}$ are interpreted as column vectors and $\left(\cdot\right)^{\mathrm{T}}$ denotes the vector transpose operator. The set of positive integers excluding 0 is denoted by $\N$. For $a\in\R,$ $\R_{\geq a}$ denotes the interval $\left[a,\infty\right)$ and $\R_{>a}$ denotes the interval $\left(a,\infty\right)$. A set-valued map from $A$ to the subsets of $B$ is denoted by $F:A\rightrightarrows B$. For a set $A$, the convex hull, the closed convex hull, the closure, the interior, and the boundary are denoted by $\co A$, $\overline{\co}A$, $\overline{A}$, $\mathring{A}$, and $\mathrm{bd}\left(A\right)$, respectively. If $a\in\R^{m}$ and $b\in\R^{n}$ then $\left[a \,\, ; \,\, b\right]$ denotes the concatenated vector $\begin{bmatrix}a\\b\end{bmatrix}\in\R^{m+n}$. For $A\subseteq\R^{m}$, $B\subseteq\R^{n}$, the set $\left\{ \left[a \,\, ; \,\, b\right]\mid a\in A,b\in B\right\} $ is denoted by $\begin{bmatrix}A\\B\end{bmatrix}$ or $\left[A \,\, ; \,\,  B\right]$. For $A,B\subseteq\R^{n}$, $A^{\mathrm{T}}B$ denotes the set $\left\{ a^{\mathrm{T}}b\mid a\in A,b\in B\right\} $, $A\pm B$ denotes the set $\left\{ a\pm b\in\R^{n}\mid a\in A,b\in B\right\} $, and $A\left(\geq\right)\leq B$ implies $\left\Vert a\right\Vert\left(\geq\right)\leq\left\Vert b\right\Vert$, $\forall a\in A$, and $\forall b\in B$. For $x\in\R^{n}$ and $r,l>0$, the sets $\left\{ y\in\R^{n}\mid\left\Vert x-y\right\Vert \leq r\right\} $, $\left\{ y\in\R^{n}\mid\left\Vert x-y\right\Vert <r\right\} $, and $\left\{ y\in\R^{n}\mid r\leq\left\Vert y\right\Vert\leq l\right\} $ are denoted by $\overline{\B}\left(x,r\right)$, $\B\left(x,r\right)$ and $\mathrm{D}\left(r,l\right)$, respectively. If $a\in\R$ then $\left|a\right|$ denotes the absolute value and if $A$ is a set then $\left|A\right|$ denotes its cardinality. For $ A\subset \R^{n} $ and $ x\in \R^{n} $, $ \dist\left(x,A\right)\coloneqq\inf_{y\in A}\left\Vert x-y \right\Vert $. Essentially bounded, $n-$times continuously differentiable, and locally Lipschitz functions with domain $A$ and codomain $B$ are denoted by $\mathcal{L}_{\infty}\left(A,B\right)$, $\mathcal{C}^{n}\left(A,B\right)$, and $\lip\left(A,B\right)$, respectively. The zero element of $\R^{n}$ is denoted by $0_{n}$, with the subscript $n$ suppressed whenever clear from the context. The notation $ \dot{V} $ is reserved for the total derivative of $ V $ with respect to time.

\section{Differential inclusions}\label{sec:diff inclu}

Let $F:\Omega\rightrightarrows\mathcal{\R}^{n}$ be a set-valued map. Consider the differential inclusion
\begin{equation}
	\dot{x}\in F\left(x,t\right). \label{eq:Inclusion}
\end{equation}
A locally absolutely continuous function $ x:\mathcal{I}_{x}\to\mathcal{D} $ is called a \emph{solution} to (\ref{eq:Inclusion}), with \emph{interval of existence} $ \mathcal{I}_{x} = \ropen{t_{0},T} $, for some $ 0 \leq t_{0} < T \leq \infty $, if $\dot{x}\left(t\right)\in F\left(x\left(t\right),t\right)$, for almost all $t\in\mathcal{I}_{x}$ \cite[p. 50]{SCC.Filippov1988}. A solution is called \emph{complete} if $\mathcal{I}_{x}=\R_{\geq t_{0}}$ and \emph{maximal} if it does not have a proper right extension\footnote{A solution $ y:\ropen{t_{0},T_{y}}\to\R^{n} $ to (\ref{eq:Inclusion}) is a \emph{(proper) right extension} of a solution $ x:\ropen{t_{0},T_{x}}\to\R^{n} $ to (\ref{eq:Inclusion}) if $ T_{y}\left(>\right)\geq T_{x} $ and $ y\left(t\right)=x\left(t\right),\forall t\in \ropen{t_{0},T_{x}} $.} which is also a solution to (\ref{eq:Inclusion}). If a solution is maximal and if the set $\overline{\left\{ x\left(t\right)\mid t\in\mathcal{I}_{x}\right\} }$ is compact, then the solution is called \emph{precompact}. Similar to \cite[Prop. 1]{SCC.Ryan1990}, Zorn's lemma can be used to show that every solution to (\ref{eq:Inclusion}) admits a right extension that is also a maximal solution to (\ref{eq:Inclusion}). Let $ \mathscr{S} \left(\mathcal{E}\right) $ denote the set of all maximal solutions to (\ref{eq:Inclusion}) such that $ \left(x\left(t_{0}\right),t_{0}\right) \in \mathcal{E} \subseteq \Omega $
\begin{preprint}
	(in the case of an autonomous system, $\mathscr{S}\left(A\right)$ denotes the set of all maximal solutions to $\dot{x}\in F\left(x\right)$ where $x\left(t_{0}\right)\in A\subset\R^{n}$)
\end{preprint}
\begin{publishedArticle}
	\!\!.
\end{publishedArticle}
The discussion in this article concerns set-valued maps that define differential inclusions that admit local solutions. 
\begin{dfntn}\label{def:existence}
	Let $ F : \Omega \rightrightarrows \mathcal{\R}^{n} $ be a set-valued map and $ \mathcal{E} \subseteq \Omega $. The differential inclusion (\ref{eq:Inclusion}) is said to \emph{admit local solutions over $ \mathcal{E} $} if for all $ \left(y,t_{0}\right) \in \mathcal{E} $, there exists  $ T \in \R_{>t_{0}} $ and a locally absolutely continuous function $ x : \ropen{t_{0},T} \to \mathcal{D} $ such that $ x\left(t_{0}\right) = y $ and $ \dot{x} \left(t\right) \in F\left(x\left(t\right) , t\right)$ for almost all $ t \in \ropen{t_{0},T} $. \defnEnd
\end{dfntn}
Sufficient conditions for the existence of local solutions to differential inclusions can be found in \cite[\S7, Thm. 1]{SCC.Filippov1988} and \cite[\S7, Thm. 5]{SCC.Filippov1988}. To assert the existence of complete solutions, the following notions of invariance are utilized in this article.
\begin{dfntn}\label{def:forwardInvariance}
	A set $ A \subseteq \mathcal{D} $ is called \emph{weakly forward invariant} with respect to (\ref{eq:Inclusion}) if $ \forall x_{0} \in A $, $ \exists x \left(\cdot\right) \in \mathscr{S} \left(\left\{x_{0}\right\} \times \R_{\geq 0}\right) $ such that $ x\left(t\right) \in A$, $ \forall t \in \mathcal{I}_{x} $. It is called \emph{strongly forward invariant} with respect to (\ref{eq:Inclusion}) if every $ x \left(\cdot\right) \in \mathscr{S} \left(A \times \R_{\geq 0}\right) $ satisfies $x \left(t\right) \in A $, $\forall t \in \mathcal{I}_{x} $. \defnEnd
\end{dfntn}
Forward invariance of a set $ A \subseteq \mathcal{D} $ in the sense of Def. \ref{def:forwardInvariance} does not imply completeness of any $ x\left(\cdot\right) \in \mathscr{S} \left(A \times \R_{\geq 0}\right) $ since $ x\left(\cdot\right) $ can exit $ \mathcal{D} $ in finite time, resulting in a finite interval of existence $ \mathcal{I}_{x} $. However, the following Lemma, which is a slight generalization of \cite[Prop. 2]{SCC.Ryan1990}, implies that under general conditions on $ F $, if $ A $ is also compact then $ \mathscr{S} \left(A \times \R_{\geq 0}\right) $ contains complete solutions, and under strong forward invariance of $ A $, all solutions in $ \mathscr{S} \left(A \times \R_{\geq 0}\right) $ are complete.
\begin{lmm}\label{lem:Precompact implies complete}
	Let $F:\Omega\rightrightarrows\mathcal{\R}^{n}$ be a set-valued map such that \eqref{eq:Inclusion} admits local solutions over $\Omega$. Let $x\left(\cdot\right)$ be a maximal solution to (\ref{eq:Inclusion}) such that $\overline{\left\{ x\left(t\right)\mid t\in\mathcal{I}_{x}\right\} }\subset\mathcal{D}$. If the set $\cup_{t\in\mathcal{\mathcal{J}}}F\left(x\left(t\right),t\right)$ is bounded for every subinterval $\mathcal{J}\subseteq\mathcal{I}_{x}$ of finite length, then $x\left(\cdot\right)$ is complete.
\end{lmm}
\begin{proof}
	For the sake of contradiction, assume that the interval of existence, $ \mathcal{I}_{x} $, is finite. That is, $\mathcal{I}_{x}=\left[t_{0},T\right)$ for some $t_{0}<T<\infty$. Boundedness of the set $\cup_{t\in\ropen{t_{0},T}}F\left(x\left(t\right),t\right)$ implies that $\dot{x}\left(\cdot\right)\in\mathcal{L}_{\infty}\left(\left[t_{0},T\right),\R^{n}\right)$. Since $x\left(\cdot\right)$ is locally absolutely continuous on $ \ropen{t_{0},T} $, it can be concluded that $\forall t_{1},t_{2}\in\left[t_{0},T\right)$, $\left\Vert x\left(t_{2}\right)-x\left(t_{1}\right)\right\Vert _{2}=\left\Vert \int_{t_{1}}^{t_{2}}\dot{x}\left(\tau\right)\diff\tau\right\Vert _{2}$. Furthermore, $\dot{x}\left(\cdot\right)\in\mathcal{L}_{\infty}\left(\left[t_{0},T\right),\R^{n}\right)$ implies that $\left\Vert \int_{t_{1}}^{t_{2}}\dot{x}\left(\tau\right)\diff\tau\right\Vert _{2}\leq\int_{t_{1}}^{t_{2}}M\diff\tau$, where $M$ is a positive constant. Thus, $\left\Vert x\left(t_{2}\right)-x\left(t_{1}\right)\right\Vert _{2}\leq M\left|t_{2}-t_{1}\right|$, and hence, $x\left(\cdot\right)$ is uniformly continuous on $\left[t_{0},T\right)$. Therefore, $x\left(\cdot\right)$ admits a continuous extension $x^{\prime}:\left[t_{0},T\right]\to\R^{n}$ \cite[Chapter 4, Exercise 13]{SCC.Rudin1976}. Since $x^{\prime}\left(\cdot\right)$ is continuous, $ \mathcal{D} $ is open, and $\overline{\left\{ x\left(t\right)\mid t\in\left[t_{0},T\right)\right\} }\subset\mathcal{D}$, it is clear that $x^{\prime}\left(T\right)\in\mathcal{D}$. Since \eqref{eq:Inclusion} admits local solutions over $\Omega$, $x^{\prime}\left(\cdot\right)$ can be extended into a solution to (\ref{eq:Inclusion}) on the interval $\ropen{t_{0},T^{\prime}}$ for some $T^{\prime}>T$, which contradicts the maximality of $x\left(\cdot\right)$. Hence, $x\left(\cdot\right)$ is complete.
\end{proof}
\begin{rmrk}
	The hypothesis of Lemma \ref{lem:Precompact implies complete}, that the set $\cup_{t\in\mathcal{\mathcal{J}}}F\left(x\left(t\right),t\right)$ needs to be bounded for every subinterval $\mathcal{J}\subseteq\mathcal{I}_{x}$ of finite length, is met if, e.g., $ \left( x , t \right) \mapsto F \left( x , t \right) $ is locally bounded over $\Omega$ and $ x\left(\cdot\right) $ is precompact (cf.\cite[Prop. 5.15]{SCC.Rockafellar.Wets2009}).
\end{rmrk}

The following section presents a summary of the relevant Lyapunov methods that utilize Clarke's notion of generalized directional derivatives and gradients \cite[p. 39]{SCC.Clarke1990} for the analysis of differential inclusions. 

\section{Set-valued derivatives\label{sec:Set-valued-derivatives}}

Clarke gradients are utilized in \cite{SCC.Shevitz.Paden1994} by Shevitz and Paden to introduce the following set-valued derivative of a locally Lipschitz, positive definite (i.e., locally positive definite in the sense of \cite[Sec. 5.2, Def. 3]{SCC.Vidyasagar2002} at $ \left(x,t\right) $, for all $ \left(x,t\right)$ in its domain) candidate Lyapunov function that is regular (i.e., regular at $ \left(x,t\right) $, in the sense of \cite[Def. 2.3.4]{SCC.Clarke1990}, for all $ \left(x,t\right)$ in its domain).
\begin{dfntn}\label{def:sastryDerivative}
	\cite{SCC.Shevitz.Paden1994} Given a regular function $V\in\lip\left(\Omega,\R\right)$, and a set-valued map $F:\Omega\rightrightarrows\mathcal{\R}^{n}$, the \emph{set-valued derivative of $V$ in the direction(s) $ F $} is defined as
	\[
		\dot{\tilde{V}}\left(x,t\right)\coloneqq\bigcap_{p\in\partial V\left(x,t\right)}p^{\mathrm{T}}\begin{bmatrix}F\left(x,t\right)\\\left\{1\right\}\end{bmatrix}, \forall \left(x,t\right)\in\Omega,
	\]
	where $\partial V$ denotes the \emph{Clarke gradient} of $V$, defined as (see also, \cite[Thm. 2.5.1]{SCC.Clarke1990})
	\begin{equation}
		\partial V  \left( x,t \right) \coloneqq 
		\overline{\co}\left\{\lim \nabla V \left( x_{i},t_{i} \right) | \left( x_{i},t_{i} \right) \rightarrow \left( x,t \right) ,\left( x_{i},t_{i} \right) \in\Omega\setminus\left(\Omega_{V}\cup S\right) \right\}, \forall \left(x,t\right)\in\Omega, \label{eq:Clarke-1}
	\end{equation}
	where $\Omega_{V}$ is the set of Lebesgue measure zero where the gradient $ \nabla V $ of $V$ is not defined and $ S\subset\Omega $ is any other set of Lebesgue measure zero.\defnEnd
\end{dfntn}
Lyapunov stability theorems developed using the set-valued derivative $\dot{\tilde{V}}$ exploit the property that every upper bound of the set $\dot{\tilde{V}}\left(x\left(t\right),t\right)$ is also an upper bound of $\dot{V}\left(x\left(t\right),t\right)$, for almost all $t$ where $\dot{V}\left(x\left(t\right),t\right)$ exists. The aforementioned fact is a consequence of the following proposition.
\begin{prpstn}\label{prop:Chain Rule}
	\cite{SCC.Shevitz.Paden1994} Let $x:\mathcal{I}_{x}\to\mathcal{D}$ be a solution to (\ref{eq:Inclusion}). If $V\in\lip\left(\Omega,\R\right)$ is a regular function, then $\dot{V}\left(x\left(t\right),t\right)$ exists for almost all $t\in\mathcal{I}_{x}$ and $\dot{V}\left(x\left(t\right),t\right)\in\dot{\tilde{V}}\left(x\left(t\right),t\right)$, for almost all $t\in\mathcal{I}_{x}$.
\end{prpstn}
\begin{proof}
	See \cite[Thm. 2.2]{SCC.Shevitz.Paden1994}.
\end{proof}
In \cite{SCC.Bacciotti.Ceragioli1999}, the notion of a set-valued derivative is further generalized via the following definition.
\begin{dfntn}\label{def:Bacciotti set valued}
	\cite{SCC.Bacciotti.Ceragioli1999} For a regular function $V\in\lip\left(\Omega,\R\right)$ and a set-valued map $F:\Omega\rightrightarrows\mathcal{\R}^{n}$, the \emph{set-valued derivative of $V$ in the direction(s) $ F $} is defined as
	\[	
		\dot{\overline{V}}  \left( x,t \right)  \coloneqq  
		\left\{ a\in\R \mid \exists q \in F \left( x,t \right)  \mid  p^{\mathrm{T}}  \left[q \,\, ; \,\, 1 \right]  = a,\forall p\in\partial V  \left( x,t \right)\right\}, \forall \left(x,t\right)\in\Omega.
		\pushedTriangle
	\]
\end{dfntn}
The set-valued derivative in Def. \ref{def:Bacciotti set valued} results in less conservative statements of Lyapunov stability than Def. \ref{def:sastryDerivative} since it is contained within the set-valued derivative in Def. \ref{def:sastryDerivative} and, as evidenced by \cite[Example 1]{SCC.Bacciotti.Ceragioli1999}, the containment can be strict. The Lyapunov stability theorems developed in \cite{SCC.Bacciotti.Ceragioli1999} exploit the property that Prop. \ref{prop:Chain Rule} also holds for $\dot{\overline{V}}$ (see \cite[Lemma 1]{SCC.Bacciotti.Ceragioli1999}).

\begin{preprint}
	In the following, notions of Lyapunov stability for differential inclusions are introduced.\footnote{While the results in this paper are stated in terms of stability at the origin, they extend in a straightforward manner to stability of arbitrary compact sets.}
	\begin{dfntn}\label{def:stability}
		The differential inclusion $\dot{x}\in F\left(x\right)$ is said to be (strongly)
		\begin{enumerate}[label=(\alph*)]
			\item \emph{stable} at $x=0$ if $\forall\epsilon>0$ $\exists\delta>0$ such that every $x\left(\cdot\right)\in\mathscr{S}\left(\overline{\B}\left(0,\delta\right)\right)$ is complete and satisfies $ x\left(t\right)\in\overline{\B}\left(0,\epsilon\right)$, $\forall t\geq0$.
			\item \emph{asymptotically stable} at $x=0$ if it is stable at $x=0$ and $\exists c>0$ such that every $x\left(\cdot\right)\in\mathscr{S}\left(\overline{\B}\left(0,c\right)\right)$ is complete and satisfies $\lim_{t\to\infty}\left\Vert x\left(t\right)\right\Vert =0$.
			\item \emph{globally asymptotically stable} at $x=0$ if it is stable at $x=0$ and every $x\left(\cdot\right)\in\mathscr{S}\left(\R^{n}\right)$ is complete and satisfies $\lim_{t\to\infty}\left\Vert x\left(t\right)\right\Vert =0$.\defnEnd
		\end{enumerate}
	\end{dfntn}
	The following proposition is an example of a typical Lyapunov stability result for time-invariant differential inclusions that utilizes set-valued derivatives of the candidate Lyapunov function. The proposition combines \cite[Thm. 2]{SCC.Bacciotti.Ceragioli1999} and a specialization of \cite[Thm. 3.1]{SCC.Shevitz.Paden1994}.
	\begin{prpstn}\label{prop:Basic Set-Valued Thm.}
		Let $F:\R^{n}\rightrightarrows\R^{n}$ be an upper semi-continuous map with compact, nonempty, and convex values. If $V\in\lip\left(\R^{n},\R\right)$ is a positive definite and regular function such that either\footnote{The definitions of $\dot{\overline{V}}$ and $\dot{\tilde{V}}$ translate to time-invariant systems as $\dot{\overline{V}}{}^{F}\left(x\right)=\left\{ a\in\R\mid\exists q\in F\left(x\right)\mid p^{\mathrm{T}}q=a,\forall p\in\partial V\left(x\right)\right\} $ and $\dot{\tilde{V}}\left(x\right)\coloneqq\bigcap_{p\in\partial V\left(x\right)}p^{\mathrm{T}}F\left(x\right)$, respectively.} $\max\dot{\overline{V}}\left(x\right)\leq0$ or $\max\dot{\tilde{V}}\left(x\right)\leq0$, $\forall x\in\R^{n}$, then $\dot{x}\in F\left(x\right)$ is stable at $x=0$.
	\end{prpstn}
	\begin{proof}
		See \cite[Thm. 2]{SCC.Bacciotti.Ceragioli1999} and \cite[Thm. 3.1]{SCC.Shevitz.Paden1994}.
	\end{proof}
\end{preprint}

Inspired by \cite{SCC.Shevitz.Paden1994} and \cite{SCC.Bacciotti.Ceragioli1999}, the following section presents a novel notion of reduced differential inclusions that results in statements of Lyapunov theorems
\begin{publishedArticle}
	\!\!
\end{publishedArticle}
\begin{preprint}
	such as Prop. \ref{prop:Basic Set-Valued Thm.}.
\end{preprint}
that are less conservative than those available in the literature.

\section{Reduced differential inclusions\label{sec:Reduced-Inclusions}}
By definition, $\dot{\overline{V}}\left(x,t\right)\subseteq\dot{\tilde{V}}\left(x,t\right)$, $ \forall \left(x,t\right)\in\Omega $, which, assuming compact values, implies $\max\dot{\overline{V}}\left(x,t\right)\leq\max\dot{\tilde{V}}\left(x,t\right)$, $ \forall \left(x,t\right)\in\Omega $. In some cases, $\max\dot{\overline{V}}$ can be strictly smaller than $\max\dot{\tilde{V}}$ and Lyapunov theorems based on $\dot{\overline{V}}$ can be less conservative than those based on $\dot{\tilde{V}}$ \cite[Example 1]{SCC.Bacciotti.Ceragioli1999}. A tighter bound on the evolution of $V$ along an orbit of (\ref{eq:Inclusion}) can be obtained by examining the following equivalent representation of $\max\dot{\overline{V}}$:\footnote{The minimization in (\ref{eq:Ceragioli-minmax}) serves to maintain consistency of notation, but is in fact, redundant.}
\begin{equation}
	\max\dot{\overline{V}}\left(x,t\right)=\min_{p\in\partial V\left(x,t\right)}\max_{q\in G_{V}^{F}\left(x,t\right)}p^{\mathrm{T}} \left[q \,\, ; \,\, 1 \right] , \label{eq:Ceragioli-minmax}
\end{equation}
where, for any regular function $U\in\lip\left(\Omega,\R\right)$, and any set-valued map $H:\Omega\rightrightarrows\R^{n}$, the reduction $G_{U}^{H}:\Omega\rightrightarrows\R^{n}$ is defined as
\begin{equation}
	G_{U}^{H}\left(x,t\right)\coloneqq\left\{ q\in H\left(x,t\right)\mid\exists a\in\R\mid p^{\mathrm{T}} \left[q \,\, ; \,\, 1 \right] =a,\forall p\in\partial U\left(x,t\right)\right\},\forall \left(x,t\right)\in\Omega. \label{eq:G}
\end{equation}

The representation in (\ref{eq:Ceragioli-minmax}), along with
\begin{publishedArticle}
	 results such as \cite[Thm. 2]{SCC.Bacciotti.Ceragioli1999},
\end{publishedArticle}
\begin{preprint}
	 Prop. \ref{prop:Basic Set-Valued Thm.},
\end{preprint}
suggest that the only directions in $F$ that affect the stability properties of solutions to (\ref{eq:Inclusion}) are those included in $G_{V}^{F}$, that is, the directions that map the Clarke gradient of $V$ to a singleton. The key observation in this paper is that \emph{the statement above remains true even when $V$ is replaced by any arbitrary locally Lipschitz, regular function $U$.} The following proposition formalizes the aforementioned observation. For clarity, the proposition is stated here for autonomous differential inclusions. The analysis of nonautonomous differential inclusions is deferred to Thm. \ref{thm:Nonautonomous}.
\begin{prpstn}\label{prop:Reduced autonomous}
	Let $F:\R^{n}\rightrightarrows\R^{n}$ be a locally bounded map with compact values such that $\dot{x}\in F\left(x\right)$ admits local solutions over $\R^{n}$. Let $V\in\lip\left(\R^{n},\R\right)$ be a positive definite and regular function and let $U\in\lip\left(\R^{n},\R\right)$ be any other regular function. If 
	\[
		\min_{p\in\partial V\left(x\right)}\max_{q\in G_{U}^{F}\left(x\right)}p^{\mathrm{T}}q\leq0,\quad\forall x\in\R^{n},
	\]
	then $\dot{x}\in F\left(x\right)$ is stable at $x=0$.
\end{prpstn}
\begin{proof}
	The proposition follows from the more general result stated in
	\begin{publishedArticle}
		 Thm. \ref{thm:Nonautonomous}
	\end{publishedArticle}
	\begin{preprint}
		 Thm. \ref{thm:Autonomous}
	\end{preprint}
	.
\end{proof}
Prop. \ref{prop:Reduced autonomous} indicates that locally Lipschitz, regular functions help discover the admissible directions in $F$. That is, from the point of view of Lyapunov stability, only the directions in $G_{U}^{F}$ are relevant, where $U$ can be an arbitrary locally Lipschitz, regular function, possibly different from the candidate Lyapunov function $V$.
\begin{preprint}
	 If $U=V$, Prop. \ref{prop:Reduced autonomous} reduces to Prop. \ref{prop:Basic Set-Valued Thm.}.
\end{preprint}

In fact, the differential inclusion $ \dot{x}\in  G_{U}^{F}\left(x,t\right)$ is, in a sense, equivalent to the differential inclusion $ \dot{x}\in F\left(x,t\right) $. To make the equivalence precise, the following definition of a reduced set-valued map is introduced.
\begin{dfntn}\label{def:U generalized time derivative}
	Let $F:\Omega\rightrightarrows\mathcal{\R}^{n}$ be a set-valued map and $\mathcal{U}\coloneqq\left\{ U_{i}\right\} _{i\in\mathcal{N}}\subset\lip\left(\Omega,\R\right) $ be a countable collection of regular functions, indexed over $ \mathcal{N}\subseteq\N $. The set-valued map $\tilde{F}_{\mathcal{U}}:\Omega\rightrightarrows\R^{n}$, defined as
	\[
		\tilde{F}_{\mathcal{U}}\left(x,t\right)\coloneqq \bigcap_{i\in\mathcal{N}}G_{U_{i}}^{F}\left(x,t\right)=\bigcap_{i\in\mathcal{N}}\left\{ q\in F\left(x,t\right)\mid\exists a\in\R\mid p^{\mathrm{T}} \left[q \,\, ; \,\, 1 \right] =a,\forall p\in\partial U_{i}\left(x,t\right)\right\}, \forall \left(x,t\right)\in\Omega,
	\]
	is called the \emph{$\mathcal{U}-$reduced set-valued map for $ F $} and the differential inclusion $ \dot{x} \in \tilde{F}_{\mathcal{U}}\left(x,t\right) $ is called the \emph{$ \mathcal{U} - $reduced differential inclusion for \eqref{eq:Inclusion}}. If $ \mathcal{U} $ is empty, then $ \tilde{F}_{\mathcal{U}} \coloneqq F $. \defnEnd
\end{dfntn}
In other words, the $\mathcal{U}-$reduced set-valued map collects all directions $ q $ in $ F $ that, through the inner product  $ p^{\mathrm{T}} \left[q \,\, ; \,\, 1 \right] $, map the Clarke gradient of all functions in $\mathcal{U}$ to a singleton. The following theorem demonstrates the key utility of the reduction in Def. \ref{def:U generalized time derivative}, i.e., the reduced differential inclusion is found to be sufficient to characterize the solutions to (\ref{eq:Inclusion}).
\begin{thrm}\label{thm:Reduction} 
	If $x:\mathcal{I}_{x}\to\mathcal{D}$ is a solution to (\ref{eq:Inclusion}), then $\dot{x}\left(t\right)\in\tilde{F}_{\mathcal{U}}\left(x\left(t\right),t\right)$ for almost all $t\in\mathcal{I}_{x}$.
\end{thrm}
\begin{proof}
	The theorem can be proved using techniques similar to \cite[Lemma 1]{SCC.Bacciotti.Ceragioli1999}.
	\begin{comment}
		 The interested reader is referred to \cite{arXivSCC.Kamalapurkar.Dixon.ea2018} for a proof.
	\end{comment}
	\begin{publishedArticle}
		 Consider the set of times $\mathcal{T}\subseteq\mathcal{I}_{x}$ where $\dot{x}\left(t\right)$ is defined, $\dot{U}_{i}\left(x\left(t\right),t\right)$ is defined $ \forall i \in \mathcal{N} $, and $\dot{x}\left(t\right)\in F\left(x\left(t\right),t\right)$. Since $x\left(\cdot\right)$ is a solution to (\ref{eq:Inclusion}), $ \mathcal{N} $ is countable, and $U_{i}\in\lip\left(\Omega,\R\right)$, it can be concluded that $t\mapsto U_{i}\left(x\left(t\right),t\right)$ is absolutely continuous, and hence, $\mu\left(\mathcal{I}_{x}\setminus\mathcal{T}\right)=0$. The objective is to show that $\dot{x}\left(t\right) $ belongs to $\in\tilde{F}_{\mathcal{U}}\left(x\left(t\right),t\right)$ on $\mathcal{T}$, not just $F\left(x\left(t\right),t\right)$. 
		
		Since each function $U_{i}$ is locally Lipschitz, for $t\in\mathcal{T}$ the time derivative of $U_{i}$ can be expressed as
		\[
			\dot{U}_{i}\left(x\left(t\right),t\right)=\lim_{h\to0}\frac{\left(U_{i}\left(x\left(t\right)+h\dot{x}\left(t\right),t+h\right)-U_{i}\left(x\left(t\right),t\right)\right)}{h}.
		\]
		Since each $U_{i}$ is regular, for $i\geq1$,
		\begin{gather*}
			\dot{U}_{i}\left(x\left(t\right),t\right)\,\,=\,\,U_{i+}^{\prime}\left(\left[x\left(t\right) \,\, ; \,\, t\right],\left[\dot{x}\left(t\right) \,\, ; \,\, 1\right]\right)\,\,=\,\, U_{i}^{o}\left(\left[x\left(t\right) \,\, ; \,\, t\right],\left[\dot{x}\left(t\right) \,\, ; \,\, 1\right]\right)\,\, =\max_{p\in\partial U_{i}\left(x\left(t\right),t\right)}p^{\mathrm{T}}\left[\dot{x}\left(t\right) \,\, ; \,\, 1\right],\\
			\dot{U}_{i}\left(x\left(t\right),t\right)\,\,=\,\,U_{i-}^{\prime}\left(\left[x\left(t\right) \,\, ; \,\, t\right],\left[\dot{x}\left(t\right) \,\, ; \,\, 1\right]\right)\,\,=\,\, U_{i}^{o}\left(\left[x\left(t\right) \,\, ; \,\, t\right],\left[\dot{x}\left(t\right) \,\, ; \,\, 1\right]\right)\,\, =\min_{p\in\partial U_{i}\left(x\left(t\right),t\right)}p^{\mathrm{T}}\left[\dot{x}\left(t\right) \,\, ; \,\, 1\right],
		\end{gather*}
		where $U_{+}^{\prime}\left(x,v\right)\coloneqq\lim_{h\downarrow0}\frac{U\left(x+hv\right)-U\left(x\right)}{h}$ and $U_{-}^{\prime}\left(x,v\right)\coloneqq\lim_{h\uparrow0}\frac{U\left(x+hv\right)-U\left(x\right)}{h}$ denote the right and left directional derivatives, and $U^{o}\left(x,v\right)\coloneqq\limsup_{y\to x,h\downarrow0}\frac{U\left(y+hv\right)-U\left(y\right)}{h}$ denotes the Clarke-generalized derivative of $U$. Thus, $p^{\mathrm{T}}\left[\dot{x}\left(t\right) \,\, ; \,\, 1\right]=\dot{U}_{i}\left(x\left(t\right),t\right),\forall p\in\partial U_{i}\left(x\left(t\right),t\right)$, which implies $\dot{x}\left(t\right)\in G_{U_{i}}^{F}\left(x\left(t\right),t\right)$, for each $i$. Therefore, $\dot{x}\left(t\right)\in\tilde{F}_{\mathcal{U}}\left(x\left(t\right),t\right)$, $\forall t\in\mathcal{T}$. Since $\mu\left(\mathcal{I}_{x}\setminus\mathcal{T}\right)=0$, $\dot{x}\left(t\right)\in\tilde{F}_{\mathcal{U}}\left(x\left(t\right),t\right)$, for almost all $t\in\mathcal{I}_{x}$.
	\end{publishedArticle}
\end{proof}
Although not directly related to the current discussion, it is worth mentioning that Thm. \ref{thm:Reduction} also expands the class of differential inclusions that admit solutions, as detailed in the following corollary.
\begin{crllr}
	A differential inclusion $ \dot{x}\in G\left(x,t\right) $, with $ G:\Omega\rightrightarrows\R^{n} $, admits local solutions over $ \mathcal{E} \subseteq \Omega $ if there exists: a set-valued map, $ F:\Omega\rightrightarrows\R^{n} $, such that \eqref{eq:Inclusion} admits local solutions over $\mathcal{E}$; and a countable collection, $ \mathcal{U}\subset\lip\left(\Omega,\R\right) $, of regular functions, such that $ G $ is the $ \mathcal{U} - $reduced set-valued map for $ F $.
\end{crllr}
The following example illustrates the utility of Thm. \ref{thm:Reduction}.
\begin{xmpl}
	Consider the differential inclusion in \eqref{eq:Inclusion}, where $x\in\R$, and $F:\R\times\R_{\geq 0}\rightrightarrows\R$ is defined as
	\[
		F\left(x,t\right)\coloneqq\begin{cases}2\sgn\left(x-1\right)&\left|x\right|\neq 1,\\\left[-2,5\right]&\left|x\right|=1,\end{cases}
	\]
	where $\sgn\left(x\right)$ denotes the sign of $x$. The function $U:\R\times\R_{\geq 0}\to\R$, defined as
	\[
		U\left(x,t\right)\coloneqq\begin{cases}\left|x\right| & \left|x\right|\leq 1, \\ 2\left|x\right|-1 & \left|x\right|>1,\end{cases}
	\]
	satisfies $U\in\lip\left(\R\times\R_{\geq 0},\R\right)$. In addition, since $ U $ is convex, it is also regular \cite[Prop. 2.3.6]{SCC.Clarke1990}. The Clarke gradient of $U$ is given by
	\[
		\partial U\left(x,t\right)=\begin{cases}
			\left[\left[1,2\right] \,\, ; \,\, \left\{ 0\right\}\right]  & x=1,\\
			\left[\left[-2,-1\right] \,\, ; \,\, \left\{ 0\right\}\right]  & x=-1,\\
			\left[\left\{ \sgn\left(x\right)\right\}  \,\, ; \,\, \left\{ 0\right\}\right]  & 0<\left|x\right|<1,\\
			\left[\left\{ 2\sgn\left(x\right)\right\} \,\, ; \,\, \left\{ 0\right\}\right]  & \left|x\right|>1,\\
			\left[\left[-1,1\right] \,\, ; \,\, \left\{ 0\right\}\right] & x=0.
		\end{cases}
	\]
	The set $G_{U}^{F}$ is then given by 
	\[
		G_{U}^{F}\left(x,t\right)=\begin{cases}
			\left\{ 0\right\} & \left|x\right|=1,\\
			\emptyset & x=0,\\
			F\left(x,t\right) & \textnormal{otherwise}.
		\end{cases}
	\]
	Thm. \ref{thm:Reduction} can then be invoked to conclude that every solution $ x:\mathcal{I}_{x}\to\R $ to \eqref{eq:Inclusion} satisfies $\dot{x}\left(t\right)\in\tilde{F}_{\left\{ U\right\} }\left(x\left(t\right),t\right)=G_{U}^{F}\left(x\left(t\right),t\right)$, for almost all $t\in\mathcal{I}_{x}$.\defnEnd
\end{xmpl}

\section{Generalized time derivatives\label{sec:generalized-time-derivatives}}

Given a countable collection $ \mathcal{U}\subset\lip\left(\Omega,\R\right) $ of regular functions and a set-valued map $F:\Omega\rightrightarrows\mathcal{\R}^{n}$ with compact values, Prop. \ref{prop:Reduced autonomous} and Thm. \ref{thm:Reduction} suggest the following notion of a generalized derivative of $V$ in the direction(s) $ F $.
\begin{dfntn}\label{def:The-generalized-time}
	The \emph{$\mathcal{U}-$generalized derivative} of $V\in\lip\left(\Omega,\R\right)$ in the direction(s) $ F $, denoted by $\dot{\overline{V}}_{\mathcal{U}}:\Omega\to\R$ is defined, $\forall\left(x,t\right)\in\Omega$, as
	\begin{equation}
		\dot{\overline{V}}_{\mathcal{U}}\left(x,t\right)\coloneqq\min_{p\in\partial V\left(x,t\right)}\max_{q\in\tilde{F}_{\mathcal{U}}\left(x,t\right)}p^{\mathrm{T}} \left[q \,\, ; \,\, 1 \right] ,\label{eq:Regular}
	\end{equation}
	if $V$ is regular, and 
	\begin{equation}
		\dot{\overline{V}}_{\mathcal{U}}\left(x,t\right)\coloneqq\max_{p\in\partial V\left(x,t\right)}\max_{q\in\tilde{F}_{\mathcal{U}}\left(x,t\right)}p^{\mathrm{T}} \left[q \,\, ; \,\, 1 \right] ,\label{eq:Nonregular}
	\end{equation}
	if $V$ is not regular. The $\mathcal{U}-$generalized derivative is understood to be $-\infty$ when $\tilde{F}_{\mathcal{U}}\left(x,t\right)$ is empty.\defnEnd
\end{dfntn}
Def. \ref{def:The-generalized-time} facilitates a unified treatment of Lyapunov stability theory using regular as well as nonregular candidate Lyapunov functions. A candidate Lyapunov function will be called a Lyapunov function if the $\mathcal{U}-$generalized derivative is negative.
\begin{dfntn}
	If $V\in\lip\left(\Omega,\R\right)$ is positive definite and if $\dot{\overline{V}}_{\mathcal{U}}\left(x,t\right)\leq0,\forall\left(x,t\right)\in\Omega$, then $V$ is called a \emph{$\mathcal{U}-$generalized Lyapunov function} for (\ref{eq:Inclusion}).\defnEnd
\end{dfntn}
If $ V $ is regular, then it can be assumed, without loss of generality, that $V\in \mathcal{U}$. In this case, $\tilde{F}_{\mathcal{U}}\subseteq G_{V}^{F}$, and hence, $\dot{\overline{V}}_{\mathcal{U}}\left(x,t\right)\leq\max\dot{\overline{V}}\left(x,t\right),\forall\left(x,t\right)\in\Omega$. Thus, by judicious selection of the functions in $\mathcal{U}$, $\dot{\overline{V}}_{\mathcal{U}}\left(x,t\right)$ can be constructed to be less conservative than the set-valued derivatives in \cite{SCC.Shevitz.Paden1994} and \cite{SCC.Bacciotti.Ceragioli1999}. Naturally, if $\mathcal{U}=\left\{ V\right\} $ then $\dot{\overline{V}}_{\mathcal{U}}=\dot{\overline{V}}$.

In general, the $\mathcal{U}-$generalized derivative does not satisfy the chain rule as stated in Prop. \ref{prop:Chain Rule}. However, it satisfies the following \emph{weak chain rule} which turns out to be sufficient for Lyapunov-based analysis of differential inclusions.
\begin{thrm}\label{thm:Wbound}
	If $V\in\lip\left(\Omega,\R\right)$ and $ \mathscr{S}\left(\Omega\right)\neq\emptyset $, then $ \forall x\left(\cdot\right)\in \mathscr{S}\left(\Omega\right) $,
	\begin{equation}
		\dot{V}\left(x\left(t\right),t\right)\in\left(\partial V\left(x\left(t\right),t\right)\right)^{\mathrm{T}}\begin{bmatrix}\tilde{F}_{\mathcal{U}}\left(x\left(t\right),t\right)\\
		\left\{ 1\right\} 
		\end{bmatrix}, \label{eq:ChainRule}
	\end{equation}
	for almost all $t\in\mathcal{I}_{x}$. In addition, if there exists a function $W:\Omega\to\R$ such that $\dot{\overline{V}}_{\mathcal{U}}\left(x,t\right)\leq W\left(x,t\right)$, $\forall\left(x,t\right)\in\Omega,$ then $\dot{V}\left(x\left(t\right),t\right)\leq W\left(x\left(t\right),t\right),$
	for almost all $t\in\mathcal{I}_{x}$.
\end{thrm}
\begin{proof}
	Let $x\left(\cdot\right)\in \mathscr{S}\left(\Omega\right)$. Consider a set of times $\mathcal{T}\subseteq\mathcal{I}_{x}$ where $\dot{x}\left(t\right)$, $\dot{V}\left(x\left(t\right),t\right)$, and $\dot{U}_{i}\left(x\left(t\right),t\right)$ are defined $\forall i\geq0$ and $\dot{x}\left(t\right)\in\tilde{F}_{\mathcal{U}}\left(x\left(t\right),t\right)$. Using Thm. \ref{thm:Reduction} and the facts that $x\left(\cdot\right)$ is absolutely continuous and $V$ is locally Lipschitz, it can be concluded that $\mu\left(\mathcal{I}_{x}\setminus\mathcal{T}\right)=0$.
	
	If $V$ is regular, then arguments similar to the proof of Thm. \ref{thm:Reduction} can be used to conclude that $\dot{V}\left(x\left(t\right),t\right)=p^{\mathrm{T}}\left[\dot{x}\left(t\right) \,\, ; \,\, 1\right],\forall p\in\partial V\left(x\left(t\right),t\right),\forall t\in\mathcal{T}$. Thus, (\ref{eq:Regular}) and Thm. \ref{thm:Reduction} imply that $\dot{V}\left(x\left(t\right),t\right)\in\left(\partial V\left(x\left(t\right),t\right)\right)^{\mathrm{T}}\left[\tilde{F}_{\mathcal{U}}\left(x\left(t\right),t\right) \,\, ; \,\, \left\{1\right\} \right]$ and $\dot{V}\left(x\left(t\right),t\right)\leq W\left(x\left(t\right),t\right)$, for almost all $t\in\mathcal{I}_{x}$.
	
	If $V$ is not regular, then \cite[Prop. 4]{SCC.Ceragioli1999} (see also, \cite[Thm. 2]{SCC.Moreau.Valadier1987}) can be used to conclude that, for almost every $t\in\mathcal{I}_{x}$, $\exists p_{0}\in\partial V\left(x\left(t\right),t\right)$ such that $\dot{V}\left(x\left(t\right),t\right)=p_{0}^{\mathrm{T}}\left[\dot{x}\left(t\right) \,\, ; \,\, 1\right]$. Thus, (\ref{eq:Nonregular}) and Thm. \ref{thm:Reduction} imply that $\dot{V}\left(x\left(t\right),t\right)\in\left(\partial V\left(x\left(t\right),t\right)\right)^{\mathrm{T}}\left[\tilde{F}_{\mathcal{U}}\left(x\left(t\right),t\right) \,\, ; \,\, \left\{1\right\} \right]$ and $\dot{V}\left(x\left(t\right),t\right)\leq W\left(x\left(t\right),t\right)$ for almost all $t\in\mathcal{I}_{x}$.
\end{proof}

The following sections develop relaxed Lyapunov-like stability theorems for differential inclusions based on the properties of the $\mathcal{U}-$generalized derivative hitherto established. 

\begin{preprint}
	\section{Stability of autonomous systems\label{sec:Lyapunov-stability-theory}}
	
	In this section, $\mathcal{U}-$generalized Lyapunov functions are utilized to formulate less conservative extensions to stability and invariance results for autonomous differential inclusions of the form 
	\begin{equation}
		\dot{x}\in F\left(x\right), \label{eq:Autonomous Inclusion}
	\end{equation}
	where $F:\mathcal{D}\rightrightarrows\R^{n}$ is a set-valued map. 
	
	\subsection{Lyapunov stability}
	
	The following Lyapunov stability theorem is a consequence of Thm. \ref{thm:Wbound}.
	\begin{thrm}\label{thm:Autonomous}
		Let $0\in\mathcal{D}$ and let $F:\mathcal{D}\rightrightarrows\R^{n}$ be a locally bounded set-valued map with compact values such that \eqref{eq:Autonomous Inclusion} admits local solutions over $\mathcal{D}$. If there exists a countable collection, $ \mathcal{U}\subset\lip\left(\Omega,\R\right) $, of regular functions and a $\mathcal{U}-$generalized Lyapunov function $V:\mathcal{D}\to\R$ for (\ref{eq:Autonomous Inclusion}), then (\ref{eq:Autonomous Inclusion}) is stable at $x=0$. In addition, if $\dot{\overline{V}}_{\mathcal{U}}\left(x\right)\leq-W\left(x\right),\forall x\in\mathcal{D}$, for some positive definite function $W\in\mathcal{C}^{0}\left(\mathcal{D},\R\right)$, then (\ref{eq:Autonomous Inclusion}) is asymptotically stable at $x=0$. Furthermore, if $\mathcal{D}=\R^{n}$ and if the sublevel sets $L_{l}\coloneqq\left\{ x\in\R^{n}\mid V\left(x\right)\leq l\right\} $ are compact for all $l\in\R_{>0}$, then (\ref{eq:Autonomous Inclusion}) is globally asymptotically stable at $x=0$.
	\end{thrm}
	\begin{proof}
		Given $\epsilon>0$, let $r>0$ be such that $\overline{\B}\left(0,r\right)\subset\mathcal{D}$ and $r\in\lopen{0,\epsilon}$. Let $\beta\in\ropen{0,\min_{\left\Vert x\right\Vert =r}V\left(x\right)}$ and $L_{\beta}\coloneqq\left\{ x\in\overline{\B}\left(0,r\right)\mid V\left(x\right)\leq\beta\right\} $. Since $V$ is continuous, $\exists\delta>0$ such that $\overline{\B}\left(0,\delta\right)\subset L_{\beta}$. Let $x\left(\cdot\right)\in\mathscr{S}\left(\mathcal{D}\right)$ be a solution to (\ref{eq:Autonomous Inclusion}). Using Thm. \ref{thm:Wbound}, which implies that $t\mapsto V\left(x\left(t\right)\right)$ is nonincreasing on $\mathcal{I}_{x}$, and standard arguments (see, e.g., \cite[Thm. 4.8]{SCC.Khalil2002}), it can be shown that $L_{\beta}$ is compact, (strongly) forward invariant, and $L_{\beta}\subset\mathcal{D}$. Hence, every solution $x\left(\cdot\right)\in\mathscr{S}\left(L_{\beta}\right)$ is precompact, and by Lemma \ref{lem:Precompact implies complete}, complete. Furthermore, if $x\left(\cdot\right)\in\mathscr{S}\left(\overline{\B}\left(0,\delta\right)\right)$ then $x\left(t\right)\in\overline{\B}\left(0,\epsilon\right)$, $\forall t\in\R_{\geq t_{0}}$.
	
		In addition, if $\dot{\overline{V}}_{\mathcal{U}}\left(x\right)\leq-W\left(x\right),\forall x\in\mathcal{D}$, for some positive definite function $W\in\mathcal{C}^{0}\left(\mathcal{D},\R\right)$, then Thm. \ref{thm:Wbound} implies that $t\mapsto V\left(x\left(t\right)\right)$ is strictly decreasing on $\R_{\geq t_{0}}$ provided $x\left(t_{0}\right)\in\B\left(0,\delta\right)$. Asymptotic stability and global asymptotic stability (in the case where the sublevel sets of $V$ are compact) of (\ref{eq:Autonomous Inclusion}) at $x=0$ then follow from standard arguments (see, e.g., \cite[Section 5.3.2]{SCC.Vidyasagar2002}).
	\end{proof}
	
	The following example presents a case where tests based on $\dot{\overline{V}}$ and $\dot{\tilde{V}}$ are inconclusive but Thm. \ref{thm:Autonomous} can be used to establish asymptotic stability.
	\begin{xmpl}\label{ex:basic Example}
		Let $H:\R\rightrightarrows\R$ be defined as
		\[
			H\left(y\right) \coloneqq \begin{cases}
				\left\{0\right\}  & \left|y\right| \neq 1,\\
				\left[-1 , 1\right] & \left|y\right| = 1.
			\end{cases}
		\]
		Let $F:\R^{2} \rightrightarrows \R^{2}$ be defined as
		\[
			F\left(x\right) \coloneqq \begin{bmatrix}
				\left\{-x_{1} + x_{2}\right\} + H\left(x_{2}\right)\\
				\left\{-x_{1} - x_{2}\right\} + H\left(x_{1}\right)
			\end{bmatrix}.
		\]
		Consider the differential inclusion in \eqref{eq:Autonomous Inclusion} and the candidate Lyapunov function $ V:\R^{2} \to \R $ defined as $ V\left(x\right) \coloneqq \frac{1}{2} \left\Vert x \right\Vert_{2}^{2} $. Since $V \in \mathcal{C}^{1} \left(\R^{2},\R\right) $, the set-valued derivatives $ \dot{\overline{V}} $ in \cite{SCC.Bacciotti.Ceragioli1999} and $ \dot{\tilde{V}} $ in \cite{SCC.Shevitz.Paden1994} are bounded by
		\begin{equation}
			\dot{ \overline{V}} \left(x\right) , \dot{\tilde{V}} \left(x\right) \leq \left\{-x_{1}^{2}-x_{2}^{2}\right\} + x_{1} H\left(x_{2}\right) + x_{2} H\left(x_{1}\right), \forall x \in \R^{2}. \label{eq:Ex1Derivative}
		\end{equation}
		Since neither $\dot{\tilde{V}}$ nor $\dot{\overline{V}}$ can be shown to be negative semidefinite everywhere, the inequality in (\ref{eq:Ex1Derivative}) is insufficient to draw conclusions regarding the stability of (\ref{eq:Autonomous Inclusion}).
	
		The function $U:\R^{2} \to \R$, defined as (see Fig. \ref{fig:U})
		\begin{figure}
			\begin{center}
				\includegraphics[width=0.5\columnwidth]{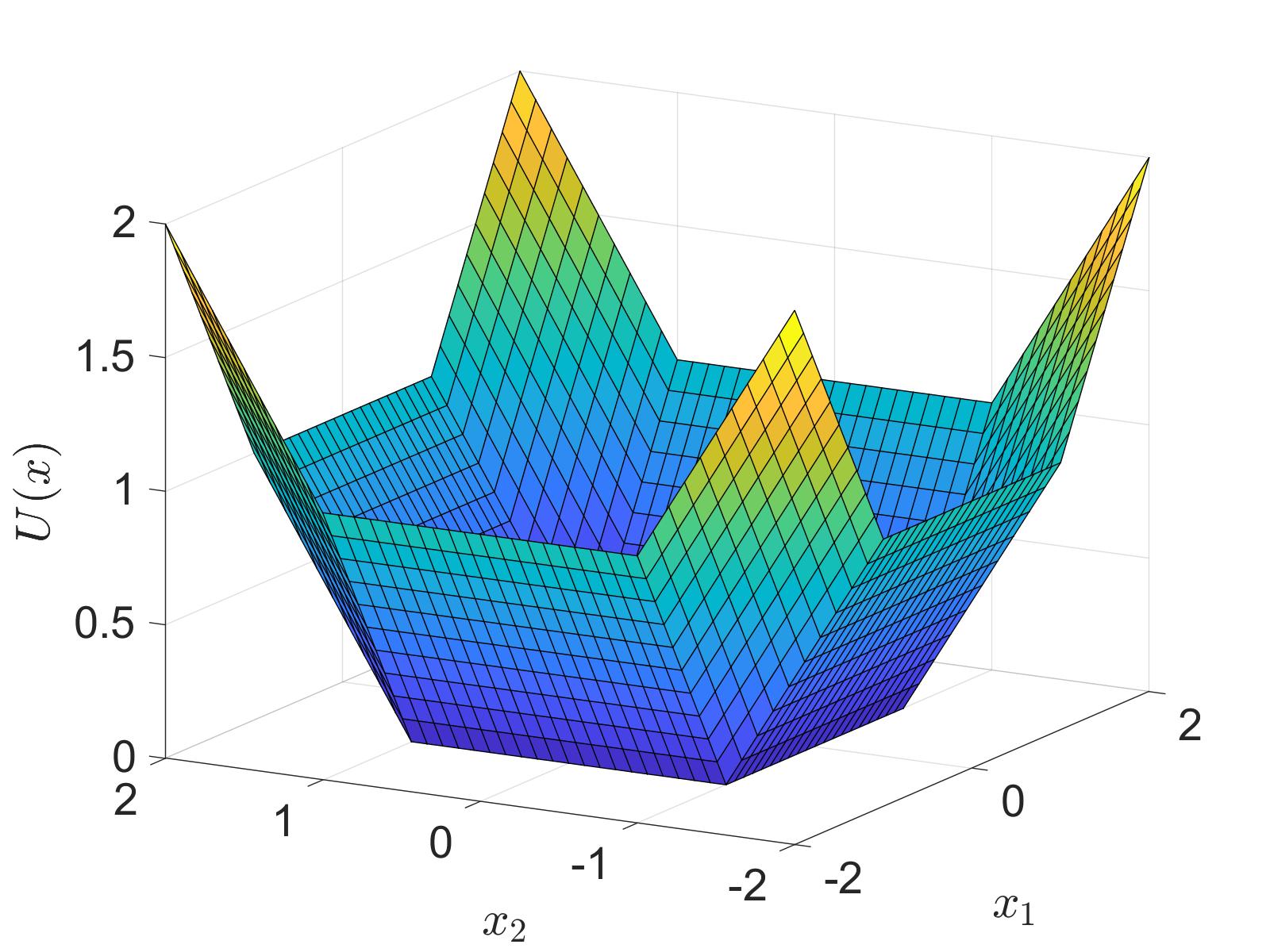}
				\caption{\label{fig:U}The function $U:\R^{2} \to \R$.}
			\end{center}
		\end{figure}
		\begin{equation*}
			U\left(x\right)\coloneqq\max\left(\left(x_{1}-1\right),0\right)-\min\left(\left(x_{1}+1\right),0\right)+\max\left(\left(x_{2}-1\right),0\right)-\min\left(\left(x_{2}+1\right),0\right), \forall x\in\R^{2},
		\end{equation*}
		satisfies $U\in\lip\left(\R^{2},\R\right)$. In addition, since $ U $ is convex, it is also regular \cite[Prop. 2.3.6]{SCC.Clarke1990}. The Clarke gradient of $U$ is given by,
		\[
			\partial U\left(x\right)=\begin{cases}
				\left[\left\{\sgn1\left(x_{1}\right)\right\} \,\, ; \,\, \left\{\sgn1\left(x_{2}\right)\right\}\right] & \left|x_{1}\right|\neq1\land\left|x_{2}\right|\neq1,\\
				\left[\overline{\co}\left\{ 0,\sgn\left(x_{1}\right)\right\} \,\, ; \,\, \left\{\sgn1\left(x_{2}\right)\right\}\right] & \left|x_{1}\right|=1\land\left|x_{2}\right|\neq1,\\
				\left[\left\{\sgn1\left(x_{1}\right)\right\} \,\, ; \,\, \overline{\co}\left\{0,\sgn\left(x_{2}\right)\right\}\right] & \left|x_{1}\right|\neq1\land\left|x_{2}\right|=1,\\
				\left[\overline{\co}\left\{0,\sgn\left(x_{1}\right)\right\} \,\, ; \,\, \overline{\co}\left\{0,\sgn\left(x_{2}\right)\right\}\right] & \left|x_{1}\right|=1\land\left|x_{2}\right|=1,
			\end{cases}
		\]
		where 
		\[
			\sgn1\left(y\right)\coloneqq\begin{cases}
				0 & -1 < y < 1,\\
				\sgn\left(y\right) & \textnormal{otherwise}.
			\end{cases}
		\]
		In this case, the reduced inclusion $G_{U}^{F}$ is given by
		\[
			G_{U}^{F}\left(x\right)=\begin{cases}
				F\left(x\right) & \left|x_{1}\right|\neq1\land\left|x_{2}\right|\neq1,\\
				\emptyset & \textnormal{otherwise}.
			\end{cases}
		\]
		Since $G_{U}^{F}\left(x\right)\subset F\left(x\right),\forall x\in\R^{2}$, $\tilde{F}_{\left\{ U\right\} }=G_{U}^{F}$. Since $V\in\mathcal{C}^{1}\left(\R^{2},\R\right)$, $\partial V\left(x\right)=\left\{ \frac{\partial V}{\partial x}\left(x\right)\right\} $, and hence, the $\left\{ U\right\} -$generalized derivative of $V$ in the direction(s) $ F $ is given by
		\begin{align*}
			\dot{\overline{V}}_{\left\{ U\right\} }\left(x\right) & =\max_{q\in\tilde{F}_{\left\{ U\right\} }\left(x\right)}\left(\frac{\partial V}{\partial x}\left(x\right)\right)^{\mathrm{T}}q, \forall x \in \R^{2}, \\
			& =\begin{cases}
				\left[x_{1} \quad x_{2}\right]\left[-x_{1}+x_{2} \,\, ; \,\, -x_{1}-x_{2}\right] & \left|x_{1}\right| \neq 1 \land \left|x_{2}\right| \neq 1,\\
				-\infty & \textnormal{otherwise},
			\end{cases}\\
			& \leq-x_{1}^{2}-x_{2}^{2}, \forall x \in \R^{2} .
		\end{align*}
		Global asymptotic stability of (\ref{eq:Autonomous Inclusion}) at $x=0$ then follows from Thm. \ref{thm:Autonomous}.\defnEnd
	\end{xmpl}
	In applications where a negative definite bound on the derivative of the candidate Lyapunov function cannot be found easily, the invariance principle is invoked. The following section develops invariance results using $\mathcal{U}-$generalized derivatives.
	
	\subsection{Invariance principle}
	
	Analogs of the Barbashin-Krasovskii-LaSalle invariance principle for autonomous differential inclusions appear in results such as \cite{SCC.Ryan1998,SCC.Bacciotti.Ceragioli1999,SCC.Alvarez.Orlov.ea2000}. Estimates of the limiting invariant set that are less conservative than those developed in \cite{SCC.Ryan1998,SCC.Bacciotti.Ceragioli1999,SCC.Alvarez.Orlov.ea2000} can be obtained by using locally Lipschitz, regular functions to reduce the admissible directions in $F$. For example, the following theorem extends the invariance principle developed by Bacciotti and Ceragioli (see \cite[Thm. 3]{SCC.Bacciotti.Ceragioli1999}). 
	\begin{thrm}\label{thm:invariance}
		Let $F:\mathcal{D}\rightrightarrows\R^{n}$ be locally bounded and outer semicontinuous \cite[Def. 5.4]{SCC.Rockafellar.Wets2009} over $\mathcal{D}$; $ F\left(x\right) $ be nonempty, convex, and compact, $ \forall x\in\mathcal{D} $; $\mathcal{U}\subset\lip\left(\mathcal{D},\R\right)$ be a countable collection of regular functions; $C\subset\mathcal{D}$ be a compact set that is strongly forward invariant with respect to (\ref{eq:Autonomous Inclusion}); $V\in\lip\left(\mathcal{D},\R\right)$; and $\dot{\overline{V}}_{\mathcal{U}}\left(x\right)\leq0$, $\forall x\in\mathcal{D}$. If $M$ is the largest weakly forward invariant set (with respect to (\ref{eq:Autonomous Inclusion})) in $\overline{E}\cap C$, where $E\coloneqq\left\{ x\in\mathcal{D}\mid\dot{\overline{V}}_{\mathcal{U}}\left(x\right)=0\right\} $, then every $x\left(\cdot\right)\in\mathscr{S}\left(C\right)$ is complete and satisfies $\lim_{t\to\infty}\dist\left(x\left(t\right),M\right)=0$.
	\end{thrm}
	\begin{proof}
		Existence of at least one $ x\left(\cdot\right)\in\mathscr{S}\left(C\right) $ follows from \cite[\S7, Thm. 1]{SCC.Filippov1988} and completeness of every $x\left(\cdot\right)\in\mathscr{S}\left(C\right)$ follows from Lemma \ref{lem:Precompact implies complete}. Given any $x\left(\cdot\right)\in\mathscr{S}\left(C\right)$, the same argument as \cite[Thm. 3]{SCC.Bacciotti.Ceragioli1999} indicates that $t\mapsto V\left(x\left(t\right)\right)$ is constant on $\omega\left(x\left(\cdot\right)\right)$, the $\omega-$limit set of $x\left(\cdot\right)$ \cite[Def. 3]{SCC.Bacciotti.Ceragioli1999}. Note that $\omega\left(x\left(\cdot\right)\right)$ is weakly invariant \cite[Prop. 2.8]{SCC.Ryan1998} (see also, \cite[\S12, Lem. 4]{SCC.Filippov1988}) and $\omega\left(x\left(\cdot\right)\right)\subset C$. Let $y:\mathcal{I}\to\omega\left(x\left(\cdot\right)\right)$ be a solution to (\ref{eq:Autonomous Inclusion}) such that $y\left(t_{0}\right)\in\omega\left(x\right)$. The existence of such a solution follows from weak invariance of $\omega\left(x\left(\cdot\right)\right)$. Since $t\mapsto V\left(x\left(t\right)\right)$ is constant on $\omega\left(x\left(\cdot\right)\right)$,  $\dot{V}\left(y\left(t\right)\right)=0,\forall t\in\R_{\geq t_{0}}$. 
	
		Let $\mathcal{T}$ be a set of time instances where $\dot{y}\left(t\right)$ is defined and $\dot{y}\left(t\right)\in\tilde{F}_{\mathcal{U}}\left(y\left(t\right)\right)$. If $V$ is regular, then arguments similar to the proof of Thm. \ref{thm:Reduction} can be used to conclude that $\forall t\in\mathcal{T}$ and $\forall p\in\partial V\left(y\left(t\right)\right)$, $0=\dot{V}\left(x\left(t\right)\right)=p^{\mathrm{T}}\dot{y}\left(t\right)$. Since $\dot{\overline{V}}_{\mathcal{U}}\left(x\right)=\min_{p\in\partial V\left(x\right)}\max_{q\in\tilde{F}_{\mathcal{U}}\left(x\right)}p^{\mathrm{T}}q\leq0,\forall x\in\mathcal{D}$ and $ p^{\mathrm{T}}\dot{y}\left(t\right)=0$, $\forall p\in\partial V\left(y\left(t\right)\right) $, it follows that $\dot{\overline{V}}_{\mathcal{U}}\left(y\left(t\right)\right)=0,\forall t\in\mathcal{T}$, which means that $y\left(t\right)\in E$, for almost all $t\in\R_{\geq t_{0}}$.
	
		If $V$ is not regular then \cite[Prop. 4]{SCC.Ceragioli1999} (see also, \cite[Thm. 2]{SCC.Moreau.Valadier1987}) can be used to conclude that for almost every $t\in\mathcal{I}$, $\exists p_{0}\in\partial V\left(y\left(t\right)\right)$ such that $0=\dot{V}\left(y\left(t\right)\right)=p_{0}^{\mathrm{T}}\dot{y}\left(t\right)$. Since, $\dot{\overline{V}}_{\mathcal{U}}\left(x\right)=\max_{p\in\partial V\left(x\right)}\max_{q\in\tilde{F}_{\mathcal{U}}\left(x\right)}p^{\mathrm{T}}q\leq0,\forall x\in\mathcal{D}$ and $p_{0}^{\mathrm{T}}\dot{y}\left(t\right)=0$ for some $ p_{0}\in \partial V\left(y\left(t\right)\right) $, it follows that $\dot{\overline{V}}_{\mathcal{U}}\left(y\left(t\right)\right)=0,\forall t\in\mathcal{T}$, which means that $y\left(t\right)\in E$ for almost all $t\in\R_{\geq t_{0}}$.
	
		Since $y\left(\cdot\right)\in\mathcal{C}^{0}\left(\R_{\geq t_{0}},\R^{n}\right)$, $y\left(t\right)\in\overline{E}$, $\forall t\in\R_{\geq t_{0}}$. That is, $\omega\left(x\left(\cdot\right)\right)\subset\overline{E}$, and hence, $\omega\left(x\left(\cdot\right)\right)\subset\overline{E}\cap C$. Since $\omega\left(x\left(\cdot\right)\right)$ is weakly invariant, $\omega\left(x\left(\cdot\right)\right)\subset M$. As a result, $\lim_{t\to\infty}\dist\left(x\left(t\right),\omega\left(x\left(\cdot\right)\right)\right)=0$ implies $\lim_{t\to\infty}\dist\left(x\left(t\right),M\right)=0$.
	\end{proof}
	
	The following corollary illustrates one of the many alternative ways to establish the existence of a compact strongly forward invariant set needed to apply Thm. \ref{thm:invariance}.
	\begin{crllr}\label{cor:invariance1}
		Let $F:\mathcal{D}\rightrightarrows\R^{n}$ be locally bounded and outer semicontinuous \cite[Def. 5.4]{SCC.Rockafellar.Wets2009} over $\mathcal{D}$; $ F\left(x\right) $ be nonempty, convex, and compact, $ \forall x\in\mathcal{D} $; and $\mathcal{U}\subset\lip\left(\mathcal{D},\R\right)$ be a countable collection of regular functions. Let $V\in\lip\left(\mathcal{D},\R\right)$ and $\dot{\overline{V}}_{\mathcal{U}}\left(x\right)\leq0$, $\forall x\in\mathcal{D}$. Let $l>0$ be such that the level set $L_{l}\coloneqq\left\{ x\in\mathcal{D}\mid V\left(x\right)\leq l\right\} $ is closed and a connected component $C_{l}$ of $L_{l}$ is bounded. If $M$ is the largest weakly forward invariant set contained in $\overline{E}\cap\overline{C_{l}}$, where $E\coloneqq\left\{ x\in\mathcal{D}\mid\dot{\overline{V}}_{\mathcal{U}}\left(x\right)=0\right\} $, then every $x\left(\cdot\right)\in\mathscr{S}\left(\overline{C_{l}}\right)$ is complete and satisfies $\lim_{t\to\infty}\dist\left(x\left(t\right),M\right)=0$. 
	\end{crllr}
	\begin{proof}
		Existence of at least one $ x\left(\cdot\right)\in\mathscr{S}\left(\overline{C_{l}}\right) $ follows from \cite[\S7, Thm. 1]{SCC.Filippov1988}. Given $ x\left(\cdot\right)\in\mathscr{S}\left(\overline{C_{l}}\right) $, it can be concluded from Thm. \ref{thm:Wbound} that $t\mapsto V\left(x\left(t\right)\right)$ is nonincreasing on $\mathcal{I}_{x}$. Note that by definition of $ \mathcal{I}_{x} $, $ x\left(t\right)\in\mathcal{D} $, for all $ t\in\mathcal{I}_{x} $. If there exists $ t_{1}\in\mathcal{I}_{x} $ such that $x\left(t_{1}\right)\notin\overline{C_{l}}$, then continuity of $x\left(\cdot\right)$, the fact that $C_{l}$ is a connected component of $L_{l}$, and the fact that $L_{l}$ is closed imply that $x\left(t_{1}\right)\notin L_{l}$, which is impossible since $t\mapsto V\left(x\left(t\right)\right)$ is nonincreasing on $\mathcal{I}_{x}$. Hence, $\overline{C_{l}}$ is strongly forward invariant on $\mathcal{I}_{x}$. Since $ C_{l} $ is bounded by assumption, $\overline{C_{l}}$ is also bounded, and as a result, $x\left(\cdot\right)$ is precompact. Therefore, by Lemma \ref{lem:Precompact implies complete}, $ x\left(\cdot\right) $ is complete. The conclusion of the corollary then follows from Thm. \ref{thm:invariance} with $C=\overline{C_{l}}$.
	\end{proof}
	
	The invariance principle is often applied to conclude asymptotic stability at the origin in the form of the following corollary.
	\begin{crllr}\label{cor:invariance}
		Let $0\in\mathcal{D}$ and $V\in\lip\left(\mathcal{D},\R\right)$ be a positive definite function. Let $F:\mathcal{D}\rightrightarrows\R^{n}$ be locally bounded and outer semicontinuous \cite[Def. 5.4]{SCC.Rockafellar.Wets2009} over $\mathcal{D}$; $ F\left(x\right) $ be nonempty, convex, and compact, $ \forall x\in\mathcal{D} $; and $\mathcal{U}\subset\lip\left(\mathcal{D},\R\right)$ be a countable collection of regular functions. If $\dot{\overline{V}}_{\mathcal{U}}\left(x\right)\leq0$, $\forall x\in\mathcal{D}$, and if, for each $\nu>0$, no complete solution to (\ref{eq:Autonomous Inclusion}) remains in the level set $\left\{ x\in\mathcal{D}\mid V\left(x\right)=\nu\right\} $, then (\ref{eq:Autonomous Inclusion}) is asymptotically stable at $x=0$. In addition, if $\mathcal{D}=\R^{n}$ and the sublevel sets $\left\{ x\in\R^{n}\mid V\left(x\right)\leq l\right\} $ are compact for all $l\in\R_{\geq0}$, then (\ref{eq:Autonomous Inclusion}) is globally asymptotically stable at $x=0$.
	\end{crllr}
	\begin{proof}
		To prove the corollary, it is first established that every complete solution converges to the origin, and then it is shown that all solutions that start in a small neighborhood of the origin are complete, and hence, converge to the origin.
		
		Given a complete solution $ x\left(\cdot\right) $, since $t\mapsto V\left(x\left(t\right)\right)$ is decreasing (by Thm. \ref{thm:Wbound}) and bounded below (because $ V $ is positive definite), $\lim_{t\to\infty}V\left(x\left(t\right)\right)=c$ for some $c\geq0$. For all $y^{*}\in\omega\left(x\left(\cdot\right)\right)$, $\exists\left\{ t_{i}\right\} _{i\in\N}\subset\R_{\geq t_{0}}$, $\lim_{i\to\infty}t_{i}=\infty$ such that $\lim_{i\to\infty}x\left(t_{i}\right)=y^{*}.$ Since $V$ is continuous, $V\left(\omega\left(x\left(\cdot\right)\right)\right)=\left\{ c\right\} $. If it can be shown that $c=0$, then positive definiteness of $V$ would imply that $\omega\left(x\left(\cdot\right)\right)=\left\{ 0\right\} $, and hence, $\lim_{t\to\infty}x\left(t\right)=0$.
	
		To prove that $c=0$ using contradiction, assume that there exists a complete solution $x:\R_{\geq t_{0}}\to\R^{n}$ such that $c>0$. Let $y:\R_{\geq t_{0}}\to\R^{n}$ be a solution to (\ref{eq:Autonomous Inclusion}) such that $y\left(t\right)\in\omega\left(x\left(\cdot\right)\right),$ $\forall t\in\R_{\geq t_{0}}.$ Such a solution exists since $\omega\left(x\left(\cdot\right)\right)$ is weakly invariant. Along the solution $y$, $V\left(y\left(t\right)\right)=c>0,$ $\forall t\in\R_{\geq t_{0}}$, which contradicts the hypothesis that no complete solution to (\ref{eq:Autonomous Inclusion}) remains in the level set $\left\{ x\in\mathcal{D}\mid V\left(x\right)=\nu\right\} $ for any $ \nu> 0 $. Therefore, $ c=0 $, and hence, every complete solution converges to the origin.
	
		Select $r>0$ such that $\overline{\B}\left(0,r\right)\subset\mathcal{D}$ and $ \delta>0 $ such that $\overline{\B}\left(0,\delta\right)\subseteq L_{\beta}\coloneqq\left\{ x\in\overline{\B}\left(0,r\right)\mid V\left(x\right)\leq\beta\right\} $, where $\beta\in\ropen{0,\min_{\left\Vert x\right\Vert =r}V\left(x\right)}$. Lyapunov stability of (\ref{eq:Autonomous Inclusion}) at $x=0$ and the fact that $L_{\beta}$ is strongly forward invariant on $ \mathcal{I}_{x} $ follows from Thm. \ref{thm:Autonomous}. Since $L_{\beta}$ is compact, all solutions starting in $L_{\beta}$ are precompact, and hence, complete, by Lemma \ref{lem:Precompact implies complete}. Since all complete solutions converge to the origin, $ x\left(\cdot\right)\in\mathscr{S}\left(\overline{\B}\left(0,\delta\right)\right) $ $ \implies $ $ \lim_{t\to\infty}x\left(t\right)=0 $; hence, (\ref{eq:Autonomous Inclusion}) is asymptotically stable at $ x=0 $. If $\mathcal{D}=\R^{n}$ and the sublevel sets $\left\{ x\in\R^{n}\mid V\left(x\right)\leq l\right\} $ are compact for all $l\in\R_{\geq0}$, then $r$, and hence, $\delta$, can be selected arbitrarily large; therefore, (\ref{eq:Autonomous Inclusion}) is globally asymptotically stable at $ x=0 $.
	\end{proof}
	
	The following example demonstrates the utility of the developed invariance principle.
	\begin{xmpl}\label{ex:invariance}
		Let $H:\R\rightrightarrows\R$ be defined as
		\[
			H\left(y\right)\coloneqq\begin{cases}
				\left\{ 0\right\}  & \left|y\right|\neq1,\\
				\left[-\frac{1}{2},\frac{1}{2}\right] & \left|y\right|=1,
			\end{cases}
		\]
		and let $F:\R^{2}\rightrightarrows\R^{2}$ be defined as 
		\begin{equation}
			F\left(x\right)\coloneqq\begin{bmatrix}\left\{ x_{2}\right\} +H\left(x_{2}\right)\\
			\left\{ -x_{1}-x_{2}\right\} +H\left(x_{1}\right)\end{bmatrix}, \forall x \in \R^2, \label{eq:invariance esample}
		\end{equation}
		and consider the differential inclusion in \eqref{eq:Autonomous Inclusion}. The candidate Lyapunov function $V:\R^{2}\to\R$ is defined as $V\left(x\right) \coloneqq \nicefrac{1}{2} \left\Vert x \right\Vert_{2}^{2}, \forall x \in \R^2 $. Since $V\in\mathcal{C}^{1}\left(\R^{2},\R\right)$, the set-valued derivatives $\dot{\overline{V}}$ in \cite{SCC.Bacciotti.Ceragioli1999} and $\dot{\tilde{V}}$ in \cite{SCC.Shevitz.Paden1994} are bounded by 
		\begin{equation}
			\dot{\overline{V}}\left(x\right),\dot{\tilde{V}}\left(x\right)\leq\left\{ -x_{2}^{2}\right\} +x_{2}H\left(x_{1}\right)+x_{1}H\left(x_{2}\right), \forall x \in \R^2. \label{eq:Ex1Derivative-1}
		\end{equation}
		That is, neither $\dot{\tilde{V}}\left(x\right)$ nor $\dot{\overline{V}}\left(x\right)$ are negative semidefinite everywhere, and hence, the inequality in (\ref{eq:Ex1Derivative-1}) is inconclusive.
	
		Let $U:\R^{2}\to\R$ be defined as in Example \ref{ex:basic Example}. The $\left\{ U\right\} -$reduced set-valued map corresponding to $F$ is given by
		\[
			\tilde{F}_{\left\{ U\right\} }\left(x\right)=\begin{cases}
				F\left(x\right) & \left|x_{1}\right|\neq1\land\left|x_{2}\right|\neq1,\\
				\left[\left\{ 0\right\}  \,\, ; \,\, \left(\left\{ -1\right\} +\left[-\frac{1}{2},\frac{1}{2}\right]\right)\right] & x_{1}=1\land x_{2}=0,\\
				\left[\left\{ 0\right\}  \,\, ; \,\, \left(\left\{ 1\right\} +\left[-\frac{1}{2},\frac{1}{2}\right]\right)\right] & x_{1}=-1\land x_{2}=0,\\
				\emptyset & \textnormal{otherwise}.
			\end{cases}
		\]
		The $\left\{ U\right\} -$generalized derivative of $V$ in the direction(s) $ F $ is then given by
		\begin{align*}
			\dot{\overline{V}}_{\left\{ U\right\} }\left(x\right) & \coloneqq\max_{q\in\tilde{F}_{\left\{ U\right\} }\left(x\right)}\left[x_{1} \quad x_{2}\right] q, \forall x \in \R^2 \\
			& = \begin{cases}
				\left[x_{1} \quad x_{2}\right]\left[x_{2} \,\, ; \,\, -x_{1}-x_{2}\right] & \left|x_{1}\right| \neq 1 \land\left|x_{2}\right| \neq 1, \\
				\max\left[1 \quad 0\right]\left[\left\{ 0\right\} \,\, ; \,\, \left[-\frac{3}{2},-\frac{1}{2}\right]\right] & x_{1}=1\land x_{2}=0, \\
				\max\left[-1 \quad 0\right]\left[\left\{ 0\right\} \,\, ; \,\, \left[\frac{1}{2},\frac{3}{2}\right]\right] & x_{1}=-1\land x_{2}=0, \\
				-\infty & \textnormal{otherwise},
			\end{cases} \\
			& \leq-x_{2}^{2}, \forall x \in \R^2.
		\end{align*}
		In this case, the set $E$ in Corollary \ref{cor:invariance1} is given by $E=\left\{ x\in\R^{2}\mid x_{2}=0\right\} $. Since the level sets $L_{l}$ are bounded and connected, $\forall l\in\R_{\geq0}$, and since the largest invariant set contained within $\overline{E}\cap L_{l}$ is $\left\{ \left[0 \,\, ; \,\, 0\right]\right\} $, $\forall l\in\R_{\geq0}$, Corollary \ref{cor:invariance1} can be invoked to conclude that all solutions to (\ref{eq:Autonomous Inclusion}) converge to the origin.
	
		From Thm. \ref{thm:Wbound}, $\dot{V}\left(x\left(t\right),t\right)\leq-x_{2}^{2}\left(t\right)$ for almost all $t\in\R_{\geq0}$; hence, given any $\nu>0$, a trajectory of (\ref{eq:Autonomous Inclusion}) can remain on the level set $\left\{ x\in\R^{n}\mid V\left(x\right)=\nu\right\} $ if and only if $x_{2}\left(t\right)=0$ and $x_{1}\left(t\right)=\pm\sqrt{2\nu}$, for all $t\in\R_{\geq0}$. From Thm. \ref{thm:Reduction}, the state $\left[x_{1} \,\, ; \,\, x_{2}\right]$ can remain constant at $\left[\pm\sqrt{2\nu} \,\, ; \,\, 0\right]$ for all $t\in\R_{\geq0}$ only if $\left[0 \,\, ; \,\, 0\right]\in F\left(\left[\pm\sqrt{2\nu} \,\, ; \,\, 0\right]\right)$, which is not true for the inclusion in (\ref{eq:invariance esample}). Therefore, Corollary \ref{cor:invariance} can be invoked to conclude that the system is globally asymptotically stable at $x=0$.\defnEnd
	\end{xmpl}
	\section{Stability of nonautonomous systems\label{sec:Stability-of-Nonautonomous}}
\end{preprint}
\begin{publishedArticle}
	\section{Stability\label{sec:Stability-of-Nonautonomous}}
\end{publishedArticle}

In this section, $\mathcal{U}-$generalized derivatives are used to establish the following forms of uniform and asymptotic stability.
\begin{dfntn}\label{def:stability-nonautonomous}
	The differential inclusion in (\ref{eq:Inclusion}) is said to be (strongly)
	\begin{enumerate}[label=(\alph*)]
		\item \emph{uniformly stable} at $x=0$, if $\forall\epsilon>0$ $\exists\delta>0$ such that every $x\left(\cdot\right)\in \mathscr{S}\left(\overline{\B}\left(0,\delta\right)\times\R_{\geq 0}\right)$ is complete and satisfies $ x\left(t\right)\in\overline{\B}\left(0,\epsilon\right)$, $\forall t\in\R_{\geq t_{0}}$.
		\item \emph{globally uniformly stable} at $x=0$, if it is uniformly stable at $x=0$ and $\forall\epsilon>0$ $\exists\Delta>0$ such that every $x\left(\cdot\right)\in \mathscr{S}\left(\overline{\B}\left(0,\epsilon\right)\times\R_{\geq 0}\right)$ is complete and satisfies $ x\left(t\right)\in\overline{\B}\left(0,\Delta\right)$, $\forall t\in\R_{\geq t_{0}}$.
		\item \emph{uniformly asymptotically stable} at $x=0$ if it is uniformly stable at $x=0$ and $\exists c>0$ such that $\forall\epsilon>0$ $\exists T\geq0$ such that every $x\left(\cdot\right)\in \mathscr{S}\left(\overline{\B}\left(0,c\right)\times\R_{\geq 0}\right)$ is complete and satisfies $ x\left(t\right)\in\overline{\B}\left(0,\epsilon\right)$, $\forall t\in\R_{\geq t_{0}+T}$.
		\item \emph{globally uniformly asymptotically stable} at $x=0$ if it is uniformly stable at $x=0$ and $\forall c,\epsilon>0$ $\exists T\geq0$ such that every $ x\left(\cdot\right)\in \mathscr{S}\left(\overline{\B}\left(0,c\right)\times\R_{\geq 0}\right)$ is complete and satisfies $x\left(t\right)\in\overline{\B}\left(0,\epsilon\right)$, $\forall t\in\R_{\geq t_{0}+T}$.\defnEnd
	\end{enumerate}
\end{dfntn}
While the results in this section are stated in terms of stability of the state at the origin and uniformity with respect to time, they extend in a straightforward manner to partial stability and uniformity with respect to a part of the state (see, e.g., \cite[Def. 4.1]{SCC.Haddad.Chellaboina.ea2006}), and stability of arbitrary compact sets. 

\subsection{Lyapunov stability}

The following fundamental Lyapunov-based stability result demonstrates the utility of $\mathcal{U}-$generalized derivatives.
\begin{thrm}\label{thm:Nonautonomous}
	Let $ 0\in\mathcal{D} $ and let $F:\Omega\rightrightarrows\R^{n}$ be a locally bounded set-valued map with	compact values such that \eqref{eq:Inclusion} admits local solutions over $ \Omega $. If there exists a positive definite function $V\in\lip\left(\Omega,\R\right)$, a pair of positive definite functions $\overline{W}\:,\underline{W}\in\mathcal{C}^{0}\left(\mathcal{D},\R\right)$, and a countable collection $ \mathcal{U}\subset\lip\left(\Omega,\R\right) $ of regular functions, such that
	\begin{gather}
		\underline{W}\left(x\right)\leq V\left(x,t\right)\leq\overline{W}\left(x\right),\quad\forall\left(x,t\right)\in\Omega,\nonumber\\
		\dot{\overline{V}}_{\mathcal{U}}\left(x,t\right)\leq0,\quad\textnormal{ for all } x\in\mathcal{D},\textnormal{ and almost all }  t\in\R_{\geq 0},\label{eq:VBounds}
	\end{gather}
	then (\ref{eq:Inclusion}) is uniformly stable at $x=0$. In addition, if there exists a positive definite function $W\in\mathcal{C}^{0}\left(\mathcal{D},\R\right)$ such that 
	\begin{equation}
		\dot{\overline{V}}_{\mathcal{U}}\left(x,t\right)\leq-W\left(x\right), \label{eq:VDdotNegative}
	\end{equation}
	for all $ x\in\mathcal{D}$ and almost all $t\in\R_{\geq 0}$, then (\ref{eq:Inclusion}) is uniformly asymptotically stable at $x=0$. Furthermore, if $\mathcal{D}=\R^{n}$ and if the sublevel sets $\left\{ x\in\R^{n}\mid\underline{W}\left(x\right)\leq c\right\} $ are compact $\forall c\in\R_{\geq0}$, then (\ref{eq:Inclusion}) is globally uniformly asymptotically stable at $x=0$.
\end{thrm}
\begin{proof}
	Select $r>0$ such that $\overline{\B}\left(0,r\right)\subset\mathcal{D}$. Let $x\left(\cdot\right)\in \mathscr{S}\left(\varOmega_{c}\times\R_{\geq 0}\right)$ where $\varOmega_{c}\coloneqq\left\{ x\in\overline{\B}\left(0,r\right)|\overline{W}\left(x\right)\leq c\right\} $ for some $c\in\left[0,\min_{\left\Vert x\right\Vert _{2}=r}\underline{W}\left(x\right)\right)$. Using Thm. \ref{thm:Wbound} and \cite[Lemma 2]{SCC.Fischer.Kamalapurkar.ea2013}, 
	\begin{equation}
		V\left(x\left(t_{0}\right),t_{0}\right)\geq V\left(x\left(t\right),t\right),\quad\forall t\in\mathcal{I}_{x}. \label{eq:VDecreases}
	\end{equation}
	Using (\ref{eq:VDecreases}) and arguments similar to \cite[Thm. 4.8]{SCC.Khalil2002}, it can be shown that every $x\left(\cdot\right)\in \mathscr{S}\left(\varOmega_{c}\times\R_{\geq 0}\right)$ satisfies $x\left(t\right)\in\overline{\B}\left(0,r\right)$, for all $ t\in\mathcal{I}_{x} $. Therefore, all solutions $x\left(\cdot\right)\in \mathscr{S}\left(\varOmega_{c}\times\R_{\geq 0}\right)$ are precompact, and as a consequence of Lemma \ref{lem:Precompact implies complete}, complete. Since $\overline{W}$ is continuous and positive definite, $\exists\delta>0$ such that $\overline{\B}\left(0,\delta\right)\subset\varOmega_{c}$. Since $\delta$ is independent of $t_{0}$, uniform stability of (\ref{eq:Inclusion}) at $x=0$ is established. The rest of the proof is identical to \cite[Section 5.3.2]{SCC.Vidyasagar2002}, and is therefore omitted.
\end{proof}

In the following example, tests based on $\dot{\overline{V}}$ and $\dot{\tilde{V}}$ are inconclusive, but Thm. \ref{thm:Nonautonomous} can be invoked to conclude global uniform asymptotic stability of the origin.
\begin{xmpl}\label{exa:Nonaut}
	Let $H:\R\rightrightarrows\R$ be defined as
	\[
		H\left(y\right)\coloneqq\begin{cases}
			\left\{ 0\right\}  & \left|y\right|\neq1,\\
			\left[-\frac{1}{2},\frac{1}{2}\right] & \left|y\right|=1,
		\end{cases}
	\]
	and let $F:\R^{2}\times\R_{\geq 0}\rightrightarrows\R^{2}$ be defined as 
	\[
		F\left(x,t\right) \coloneqq \begin{bmatrix} \left\{-x_{1} + x_{2}\left(1 + g\left(t\right)\right)\right\} + H\left(x_{2}\right)\\
		\left\{-x_{1} - x_{2}\right\} + H\left(x_{1}\right) \end{bmatrix}, \forall \left(x,t\right) \in \R^{2} \times \R_{\geq 0},
	\]
	where $g\in\mathcal{C}^{1}\left(\R_{\geq 0},\R\right)$, $0\leq g\left(t\right)\leq1,\forall t \in \R_{\geq 0} $ and $\dot{g}\left(t\right)\leq g\left(t\right),\forall t \in \R_{\geq 0} $. Consider the differential inclusion in \eqref{eq:Inclusion} and the candidate Lyapunov function $V:\R^{2}\times\R_{\geq 0}\to\R$ defined as $V\left(x,t\right)\coloneqq x_{1}^{2}+\left(1+g\left(t\right)\right)x_{2}^{2}$. the candidate Lyapunov function satisfies $\left\Vert x\right\Vert _{2}^{2}\leq V\left(x,t\right)\leq2\left\Vert x\right\Vert _{2}^{2},\forall\left(x,t\right)\in\R^{2}\times\R_{\geq 0}$. In this case, since $V\in\mathcal{C}^{1}\left(\R^{2}\times\R_{\geq 0},\R\right)$, similar to \cite[Example 4.20]{SCC.Khalil2002}, the set-valued derivatives $\dot{\overline{V}}$ in \cite{SCC.Bacciotti.Ceragioli1999} and $\dot{\tilde{V}}$ in \cite{SCC.Shevitz.Paden1994} satisfy the bound $ \dot{\overline{V}} \left(x,t\right) , \dot{\tilde{V}} \left(x,t\right) \leq \left\{-2x_{1}^{2} - 2x_{2}^{2}\right\} + 2x_{1} H\left(x_{2}\right) + 2x_{2} h\left(t\right) H\left(x_{1}\right) $, where $ h\left(t\right) \coloneqq 1 + g\left(t\right) $ and the inequality $ 2 + 2g \left(t\right) - \dot{g} \left(t\right) \geq 2 $ is utilized. Therefore, neither $ \dot{\tilde{V}} \left(x,t\right) $ nor $ \dot{\overline{V}} \left(x,t\right) $ can be shown to be negative semidefinite everywhere.
	
	The function $ U_{1} : \R^{2} \times \R_{\geq 0} \to \R $, defined as (see Fig. \ref{fig:U1})
	\begin{figure}
		\begin{center}
			\includegraphics[width=0.5\columnwidth]{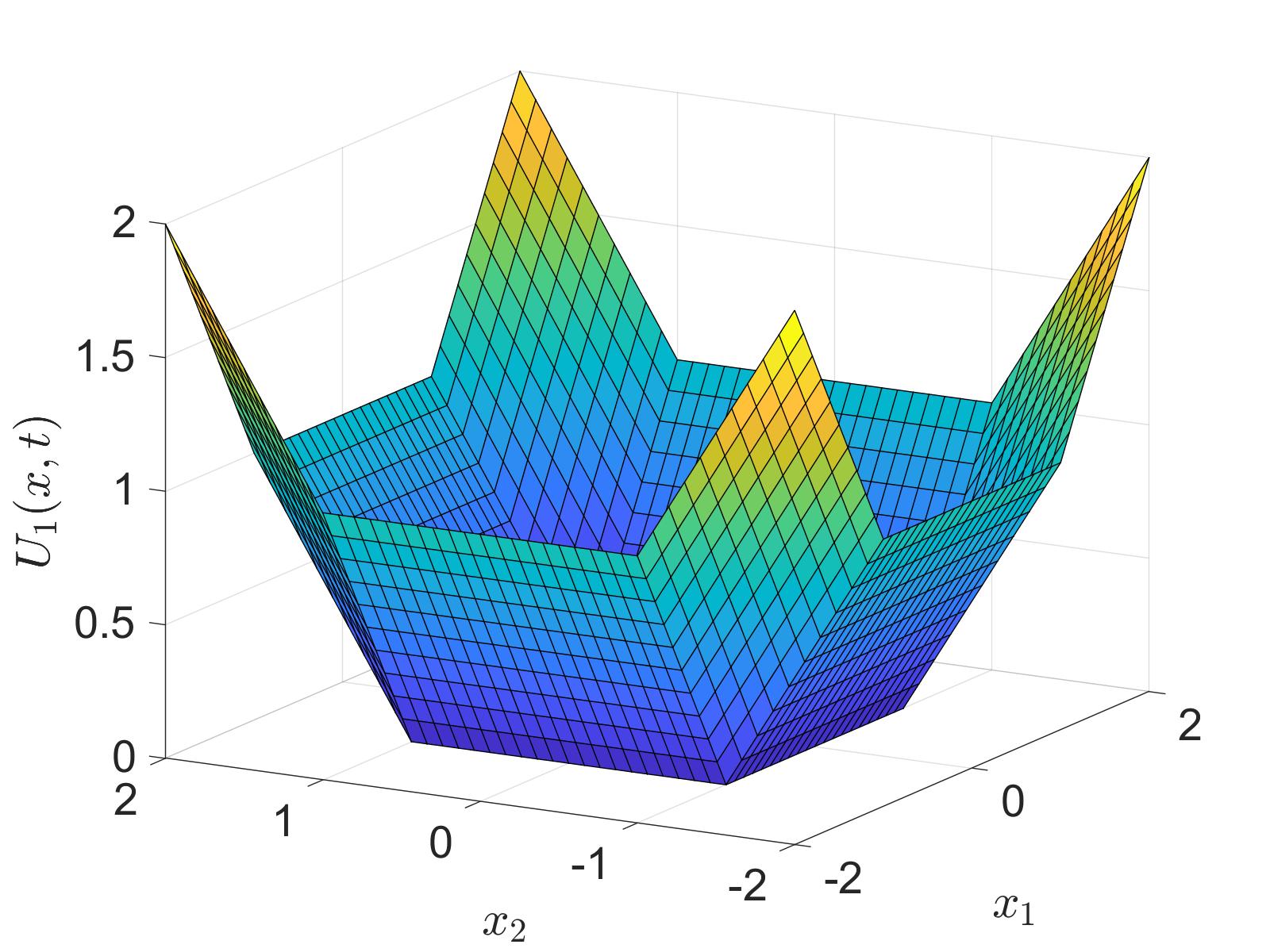}
			\caption{\label{fig:U1}A snapshot of the function $U_{1}:\R^{2}\times\R_{\geq 0} \to \R$.}
		\end{center}
	\end{figure}
	\begin{equation}
		U_{1}\left(x,t\right)=\max\left(\left(x_{1}-1\right),0\right)-\min\left(\left(x_{1}+1\right),0\right)+\max\left(\left(x_{2}-1\right),0\right)-\min\left(\left(x_{2}+1\right),0\right),\forall \left(x,t\right)\in\R^{2}\times\R_{\geq 0}, \label{eq:U}
	\end{equation}
	satisfies $U_{1}\in\lip\left(\R^{2}\times\R_{\geq 0},\R\right)$. In addition, since $ U_{1} $ is convex, it is also regular \cite[Prop. 2.3.6]{SCC.Clarke1990}. With
	\[
		\sgn1\left(y\right)\coloneqq\begin{cases}
			0 & -1<y<1,\\
			\sgn\left(y\right) & \textnormal{otherwise},
		\end{cases}
	\]
	the Clarke gradient of $U_{1}$ is given by
	\[
		\partial U_{1} \left(x,t\right) = \begin{cases}
			\left[\left\{\sgn1\left(x_{1}\right)\right\} \,\, ; \,\, \left\{\sgn1\left(x_{2}\right)\right\} \,\, ; \,\, \left\{0\right\}\right] & \left|x_{1}\right|\neq1\land\left|x_{2}\right|\neq1,\\
			\left[\overline{\co}\left\{ 0,\sgn\left(x_{1}\right)\right\} \,\, ; \,\, \left\{\sgn1\left(x_{2}\right)\right\} \,\, ; \,\, \left\{0\right\}\right] & \left|x_{1}\right|=1\land\left|x_{2}\right|\neq1,\\
			\left[\left\{\sgn1\left(x_{1}\right)\right\} \,\, ; \,\, \overline{\co}\left\{0,\sgn\left(x_{2}\right)\right\} \,\, ; \,\, \left\{0\right\}\right]  & \left|x_{1}\right|\neq1\land\left|x_{2}\right|=1,\\
			\left[\overline{\co}\left\{0,\sgn\left(x_{1}\right)\right\} \,\, ; \,\, \overline{\co}\left\{0,\sgn\left(x_{2}\right)\right\} \,\, ; \,\, \left\{ 0\right\}\right]  & \left|x_{1}\right|=1\land\left|x_{2}\right|=1,
		\end{cases}
	\]
	The $\left\{ U_{1}\right\} -$reduced set-valued map corresponding to $F$ is given by
	\[
		\tilde{F}_{\left\{ U_{1}\right\} }\left(x,t\right)=\begin{cases}
			F\left(x,t\right) & \left|x_{1}\right|\neq1\land\left|x_{2}\right|\neq1,\\
			\emptyset & \textnormal{otherwise}.
		\end{cases}
	\]
	The $\left\{ U_{1}\right\} -$generalized derivative of $V$ in the direction(s) $ F $ is then given by
	\begin{align*}
		\dot{\overline{V}}_{\left\{ U_{1}\right\} }\left(x,t\right) & \coloneqq\max_{q\in\tilde{F}_{\left\{ U_{1}\right\} }\left(x,t\right)}\left(\frac{\partial V}{\partial\left(x,t\right)}\left(x,t\right)\right)^{\mathrm{T}}\left[q \,\, ; \,\, 1\right],\forall \left(x,t\right)\in\R^{2}\times\R_{\geq 0},\\
		& =\begin{cases}
			\left[2x_{1} \quad	2x_{2}h\left(t\right) \quad \dot{g}\left(t\right)x_{2}^{2}\right]\left[-x_{1}+x_{2}h\left(t\right) \,\, ; \,\, -x_{1}-x_{2} \,\, ; \,\, 1\right] & \left|x_{1}\right| \neq 1 \land
			\left|x_{2}\right| \neq 1, \\
			-\infty & \textnormal{otherwise},
		\end{cases}\\
		& \leq-2\left\Vert x\right\Vert _{2}^{2},\forall \left(x,t\right)\in\R^{2}\times\R_{\geq 0}.
	\end{align*}
	Thm. \ref{thm:Nonautonomous} can then be invoked to conclude that \eqref{eq:Inclusion} is globally uniformly asymptotically stable at $ x=0 $.\defnEnd
\end{xmpl}

\subsection{Invariance-like results}

In applications such as adaptive control, Lyapunov methods commonly result in semidefinite Lyapunov functions (i.e., candidate Lyapunov functions with time derivatives bounded by a negative semidefinite function of the state). The following theorem establishes the fact that if the function $W$ in (\ref{eq:VDdotNegative}) is positive semidefinite then $t\mapsto W\left(x\left(t\right)\right)$ asymptotically decays to zero.
\begin{publishedArticle}
 If the differential inclusion is time-invariant, stronger results similar to LaSalle's invariance principle can also be established using $\mathcal{U}-$generalized derivatives (see \cite{arXivSCC.Kamalapurkar.Dixon.ea2018}).
\end{publishedArticle}
\begin{thrm}\label{thm:GLYT}
	Let $ 0\in \mathcal{D} $, select $r>0$ such that $\overline{\B}\left(0,r\right)\subset\mathcal{D}$, and let $F:\Omega\rightrightarrows\R^{n}$ be a set-valued map with compact values that is locally bounded, uniformly in $t$, over $\Omega$,\footnote{A set-valued map $F:\R^{n}\times\R\rightrightarrows\R^{n}$ is locally bounded, uniformly in $t$, over $ \mathcal{D}\times\mathcal{J} $ for some $ \mathcal{D}\subseteq\R^{n} $ and $ \mathcal{J}\subseteq\R $, if for every compact $K\subset\mathcal{D}$, there exists $M>0$ such that $\forall\left(x,t,y\right)$ such that $\left(x,t\right)\in K\times\mathcal{J}$, and $y\in F\left(x,t\right)$, $\left\Vert y\right\Vert _{2}\leq M$.} such that \eqref{eq:Inclusion} admits local solutions over $ \Omega $. If there exists a positive definite function $V\in\lip\left(\Omega,\R\right)$, a positive semidefinite function $W\in\mathcal{C}^{0}\left(\mathcal{D},\R\right)$, a pair of positive definite functions $\overline{W}\:,\underline{W}\in\mathcal{C}^{0}\left(\mathcal{D},\R\right)$, and a countable collection $ \mathcal{U}\subset\lip\left(\Omega,\R\right) $ of regular functions such that (\ref{eq:VBounds}) and (\ref{eq:VDdotNegative}) hold, then every solution $x\left(\cdot\right)\in \mathscr{S}\left(\varOmega_{c}\times\R_{\geq 0}\right)$, with $\varOmega_{c}\coloneqq\left\{ x\in\overline{\B}\left(0,r\right)\mid\overline{W}\left(x\right)\leq c\right\} $ and $c\in\left[0,\min_{\left\Vert x\right\Vert _{2}=r}\underline{W}\left(x\right)\right)$, is complete, bounded, and satisfies $\lim_{t\to\infty}W\left(x\left(t\right)\right)=0$.
\end{thrm}
\begin{proof}
	Similar to the proof of \cite[Corollary 1]{SCC.Fischer.Kamalapurkar.ea2013}, it is established that the bounds on $\dot{\overline{V}}_{F}$ in (\ref{eq:Regular}) and (\ref{eq:Nonregular}) imply that $ V $ is nonincreasing along all the solutions to (\ref{eq:Inclusion}). The nonincreasing property of $ V $ is used to establish boundedness of $x\left(\cdot\right)$, which is used to prove the existence and uniform continuity of complete solutions. Barb{\u{a}}lat's lemma \cite[Lemma 8.2]{SCC.Khalil2002} is then used to conclude the proof.

	Let $x\left(\cdot\right)\in \mathscr{S}\left(\varOmega_{c}\times\R_{\geq 0}\right)$. Using Thm. \ref{thm:Wbound} and \cite[Lemma 2]{SCC.Fischer.Kamalapurkar.ea2013}, $V\left(x\left(t_{0}\right),t_{0}\right)\geq V\left(x\left(t\right),t\right),\quad\forall t\in\mathcal{I}_{x}$. Arguments similar to \cite[Thm. 4.8]{SCC.Khalil2002} can then be used to show that every $x\left(\cdot\right)\in \mathscr{S}\left(\varOmega_{c}\times\R_{\geq 0}\right)$ satisfies $ x \left(t\right) \in \overline{\B} \left(0,r\right), \forall t \in \mathcal{I}_{x} $. Therefore, all solutions $x\left(\cdot\right)\in \mathscr{S}\left(\varOmega_{c}\times\R_{\geq 0}\right)$ are precompact, and as a consequence of Lemma \ref{lem:Precompact implies complete}, complete.

	To establish uniform continuity of the solutions, it is observed that since $F$ is locally bounded, uniformly in $t$, over $\Omega$, and $x\left(t\right)\in\overline{\B}\left(0,r\right)$ on $\R_{\geq t_{0}}$, the map $t\mapsto F\left(x\left(t\right),t\right)$ is uniformly bounded on $\R_{\geq t_{0}}$. Hence, $ \dot{x} \in \mathcal{L}_{\infty} \left(\R_{\geq t_{0}} , \R^{n}\right) $. Since $x\left(\cdot\right)$ is locally absolutely continuous, $\forall t_{1},t_{2}\in\R_{\geq t_{0}}$, $\left\Vert x\left(t_{2}\right)-x\left(t_{1}\right)\right\Vert _{2}=\left\Vert \int_{t_{1}}^{t_{2}}\dot{x}\left(\tau\right)\diff\tau\right\Vert _{2}$. Since $\dot{x}\in \mathcal{L}_{\infty} \left(\R_{\geq t_{0}} , \R^{n}\right) $, $\left\Vert \int_{t_{1}}^{t_{2}}\dot{x}\left(\tau\right)\diff\tau\right\Vert _{2}\leq\int_{t_{1}}^{t_{2}}M\diff\tau$, where $M$ is a positive constant. Thus, $\left\Vert x\left(t_{2}\right)-x\left(t_{1}\right)\right\Vert _{2}\leq M\left|t_{2}-t_{1}\right|$, and hence, $x\left(\cdot\right)$ is uniformly continuous on $\R_{\geq t_{0}}$.

	Since $x\mapsto W\left(x\right)$ is continuous and $\overline{\B}\left(0,r\right)$ is compact, $x\mapsto W\left(x\right)$ is uniformly continuous on $\overline{\B}\left(0,r\right)$. Hence, $t\mapsto W\left(x\left(t\right)\right)$ is uniformly continuous on $\R_{\geq t_{0}}$. Furthermore, $t\mapsto\intop_{t_{0}}^{t}W\left(x\left(\tau\right)\right)\diff\tau$ is monotonically increasing and from (\ref{eq:VDdotNegative}), $\intop_{t_{0}}^{t}W\left(x\left(\tau\right)\right)\diff\tau$ $\leq V\left(x\left(t_{0}\right),t_{0}\right)-V\left(x\left(t\right),t\right)$ $\leq V\left(x\left(t_{0}\right),t_{0}\right)$. Hence, $\lim_{t\to\infty}\intop_{t_{0}}^{t}W\left(x\left(\tau\right)\right)\diff\tau$ exists and is finite. By Barb{\u{a}}lat's Lemma \cite[Lemma 8.2]{SCC.Khalil2002}, $\lim_{t\to\infty}W\left(x\left(t\right)\right)=0$.
\end{proof}

In the following example $\dot{\overline{V}}$ and $\dot{\tilde{V}}$ do not have a negative semidefinite upper bound, but Thm. \ref{thm:GLYT} can be invoked to conclude partial stability.
\begin{xmpl}\label{exa:KLYT}
	Let $H:\R\rightrightarrows\R$ be defined as in Example \ref{exa:Nonaut} and let $F:\R^{2}\times\R_{\geq 0}\rightrightarrows\R^{2}$ be defined as 
	\[
		F\left(x,t\right)=\begin{bmatrix}\left\{ x_{2}\left(1+g\left(t\right)\right)\right\} +H\left(x_{2}\right)\\
		\left\{ -x_{1}-x_{2}\right\} +H\left(x_{1}\right)\end{bmatrix},
	\]
	where $g\in\mathcal{C}^{1}\left(\R_{\geq 0},\R\right)$, $0\leq g\left(t\right)\leq1,\forall t\in\R_{\geq 0}$ and $\dot{g}\left(t\right)\leq g\left(t\right),\forall t\in\R_{\geq 0}$. Consider the differential inclusion in \eqref{eq:Inclusion} and the candidate Lyapunov function $V:\R^{2}\times\R_{\geq 0}\to\R$ defined as $V\left(x,t\right)\coloneqq x_{1}^{2}+\left(1+g\left(t\right)\right)x_{2}^{2}$. The candidate Lyapunov function satisfies $\left\Vert x\right\Vert _{2}^{2}\leq V\left(x,t\right)\leq2\left\Vert x\right\Vert _{2}^{2},\forall\left(x,t\right)\in\R^{2}\times\R_{\geq 0}$. In this case, since $V\in\mathcal{C}^{1}\left(\R^{2}\times\R_{\geq 0},\R\right)$, the set-valued derivatives $\dot{\overline{V}}$ in \cite{SCC.Bacciotti.Ceragioli1999} and $\dot{\tilde{V}}$ in \cite{SCC.Shevitz.Paden1994} are bounded by
	\[
		\dot{\overline{V}}\left(x,t\right),\dot{\tilde{V}}\left(x,t\right)\leq\left\{-2x_{2}^{2}\right\}+2x_{2}h\left(t\right)H\left(x_{1}\right)+2x_{1}H\left(x_{2}\right),\forall\left(x,t\right)\in\R^{2}\times\R_{\geq 0},
	\]
	where $h\left(t\right)\coloneqq1+g\left(t\right)$ and the inequality $2+2g\left(t\right)-\dot{g}\left(t\right)\geq2$ is utilized. Thus, neither $\dot{\tilde{V}}$ nor $\dot{\overline{V}}$ are negative semidefinite everywhere.
	
	Let $U_{1}$ be defined as in \eqref{eq:U}. The $\left\{ U_{1}\right\} -$reduced set-valued map corresponding to $F$ is given by
	\[
		\tilde{F} _{\left\{ U_{1} \right\} } \left(x,t\right) = \begin{cases}
			F\left(x,t\right) & \left|x_{1}\right| \neq1\land\left|x_{2}\right| \neq1,\\
			\left[\left\{ 0\right\} \,\, ; \,\, \left(\left\{-1\right\} + \left[-\frac{1}{2},\frac{1}{2}\right]\right)\right] & x_{1} = 1\land x_{2} = 0,\\
			\left[\left\{ 0\right\} \,\, ; \,\, \left(\left\{1\right\} + \left[-\frac{1}{2},\frac{1}{2}\right]\right)\right] & x_{1} = -1\land x_{2} = 0,\\
			\emptyset & \textnormal{otherwise}.
		\end{cases}
	\]
	The $\left\{ U_{1}\right\} -$generalized derivative of $V$ in the direction(s) $ F $ is then given by
	\begin{align*}
		\dot{\overline{V}}_{\left\{ U_{1}\right\} }\left(x,t\right) & \coloneqq\max_{q\in\tilde{F}_{\left\{ U_{1}\right\} }\left(x,t\right)}\left(\frac{\partial V}{\partial\left(x,t\right)}\left(x,t\right)\right)^{\mathrm{T}} \left[q \,\, ; \,\, 1 \right] ,\forall\left(x,t\right)\in\R^{2}\times\R_{\geq 0},\\
		& =\begin{cases}
			\left[2x_{1} \quad 2x_{2}h\left(t\right) \quad \dot{g}\left(t\right)x_{2}^{2}\right]\left[x_{2}h\left(t\right) \,\, ; \,\, -x_{1}-x_{2} \,\, ; \,\, 1\right] & \left|x_{1}\right|\neq 1\land \left|x_{2}\right|\neq 1,\\
			\max\left[2 \quad 0 \quad 0\right]\left[\left\{0\right\} \,\, ; \,\, \left[\frac{3}{2},-\frac{1}{2}\right] \,\, ; \,\, 1\right] & x_{1}=1\land x_{2}=0,\\
			\max\left[-2 \quad 0 \quad 0\right]\left[\left\{ 0\right\} \,\, ; \,\, \left[\frac{1}{2},\frac{3}{2}\right] \,\, ; \,\, 1\right] &  x_{1}=-1\land x_{2}=0,\\
			-\infty & \textnormal{otherwise},
		\end{cases}\\
		& \leq-2x_{2}^{2},\forall\left(x,t\right)\in\R^{2}\times\R_{\geq 0}.
	\end{align*}
	Thm. \ref{thm:GLYT} can then be invoked to conclude that $t\mapsto  x_{1}\left(t\right) \in\mathcal{L}_{\infty}\left(\R_{\geq t_{0}},\R\right) $ and $ \lim_{t\to\infty}x_{2}\left(t\right)=0 $.\defnEnd
\end{xmpl}
Thm. \ref{thm:GLYT} and its counterparts are widely used in applications such as adaptive control to establish stability (but not asymptotic stability) of the state and convergence of a part of the state (e.g., tracking errors, but not parameter estimation errors) to the origin. Under certain excitation conditions, asymptotic stability (and as a result, convergence of the entire state to the origin) can be established using Matrosov theorems \cite{SCC.Matrosov1962}.

\subsection{Matrosov theorems}

In this section, a less conservative generalization of Matrosov results for uniform asymptotic stability of nonautonomous systems is developed. In particular, the nonsmooth version \cite[Thm. 1]{SCC.Teel.Nesic.ea2016} of the nested Matrosov theorem \cite[Thm. 1]{SCC.Loria.Panteley.ea2005} is generalized. The following definitions of Matrosov functions are inspired by \cite{SCC.Teel.Nesic.ea2016}. 
\begin{dfntn}
	Let $\gamma,\delta,\Delta>0$ be constants. A finite set of functions $\left\{ Y_{j}\right\} _{j=1}^{M}\subset\mathcal{C}^{0}\left(\overline{\B}\left(0_{m},\gamma\right)\times\mathrm{D}\left(\delta,\Delta\right),\R\right)$ is said to have the \emph{Matrosov property} relative to $\left(\gamma,\delta,\Delta\right)$ if $\forall j\in\left\{ 0,\cdots,M\right\} $,
	\[
		\left(\left(z,x\right) \in \overline{\B} \left(0_{m} , \gamma\right) \times \mathrm{D} \left(\delta , \Delta\right)\right) \land \left(Y_{i} \left(z , x\right) = 0, \forall i \in \left\{0,\cdots,j\right\} \right) \implies Y_{j+1} \left(z , x\right) \leq 0,
	\]
	where $Y_{0}\left(z,x\right)=0$ and $Y_{M+1}\left(z,x\right)=1$, $\forall\left(z,x\right)\in\overline{\B}\left(0_{m},\gamma\right)\times\mathrm{D}\left(\delta,\Delta\right)$.\defnEnd
\end{dfntn}
\begin{dfntn}\label{def:Matrosov}
	Let $\delta,\Delta>0$ be constants such that $\mathrm{D}\left(\delta,\Delta\right)\subset\mathcal{D}$. Let $F:\Omega\rightrightarrows\R^{n}$ be a set-valued map with compact values. The functions $\left\{ W_{j}\right\} _{j=1}^{M}\subset\lip\left(\Omega,\R\right)$ are said to be \emph{$ \mathcal{U} - $reduced Matrosov functions} for $\left(F,\delta,\Delta\right)$ if $\exists\phi:\Omega\to\R^{m}$, $\gamma>0$, and $\left\{ Y_{j}\right\} _{j=1}^{M}\subset\mathcal{C}^{0}\left(\overline{\B}\left(0_{m},\gamma\right)\times\mathrm{D}\left(\delta,\Delta\right),\R\right)$ such that:
	\begin{enumerate}[ref = {\thedfntn.\alph*},label=(\alph*)]
		\item \label{enu:Matrosov Property} the set of functions $\left\{ Y_{j}\right\} _{j=1}^{M}$ has the Matrosov property relative to $\left(\gamma,\delta,\Delta\right)$,
		\item \label{enu:boundedness}$\forall j\in\left\{ 1,\cdots,M\right\} $ and $\forall\left(x,t\right)\in\mathrm{D}\left(\delta,\Delta\right)\times\R_{\geq 0}$, $\max\left\{ \left|W_{j}\left(x,t\right)\right|,\left|\phi\left(x,t\right)\right|\right\} \leq\gamma$, and
		\item \label{enu:derivative Bound}$\forall j\in\left\{ 1,\cdots,M\right\} $ there exists a collection of regular functions $\mathcal{U}_{j}\subset\lip\left(\mathrm{D}\left(\delta,\Delta\right)\times\R_{\geq 0},\R\right)$ such that $\forall\left(x,t\right)\in\mathrm{D}\left(\delta,\Delta\right)\times\R_{\geq 0}$, $\dot{\overline{W}}_{\mathcal{U}_{j}}\left(x,t\right)\leq Y_{j}\left(\phi\left(x,t\right),x\right)$. \defnEnd
	\end{enumerate}
\end{dfntn}
The following technical Lemmas aid the proof of the Matrosov theorem.
\begin{lmm}\label{lem:Weird 1}
	Given $\delta>0$, $\exists\epsilon>0$ such that
	\[
		\left(\left(z,x\right)\in\overline{\B}\left(0_{m},\gamma\right)\times\mathrm{D}\left(\delta,\Delta\right)\right)\land\left(Y_{j}\left(z,x\right)=0,\forall j\in\left\{ 1,\cdots,M-1\right\} \right)\implies Y_{M}\left(z,x\right)\leq-\epsilon.
	\]
\end{lmm}
\begin{proof}
	See \cite[Claim 1]{SCC.Loria.Panteley.ea2005}.
\end{proof}
\begin{lmm}\label{lem:Weird 2}
	Let $l\in\left\{ 2,\cdots,M\right\} $, $\tilde{\epsilon}>0$, and $\tilde{Y}_{l}\in\mathcal{C}^{0}\left(\R^{m}\times\R^{n},\R\right)$. If
	\[
		\left(\left(z,x\right)\in\overline{\B}\left(0_{m},\gamma\right)\times\mathrm{D}\left(\delta,\Delta\right)\right)\land\left(Y_{j}\left(z,x\right)=0,\forall j\in\left\{ 1,\cdots,l-1\right\} \right)\implies\tilde{Y}_{l}\left(z,x\right)\leq-\tilde{\epsilon},
	\]
	then $\exists K_{l-1}>0$ such that 
	\[
		\left(\left(z,x\right)\in\overline{\B}\left(0_{m},\gamma\right)\times\mathrm{D}\left(\delta,\Delta\right)\right)\land\left(Y_{j}\left(z,x\right)=0,\forall j\in\left\{ 1,\cdots,l-2\right\} \right)\implies K_{l-1}Y_{l-1}\left(z,x\right)+\tilde{Y}_{l}\left(z,x\right)\leq-\frac{\tilde{\epsilon}}{2}.
	\]
\end{lmm}
\begin{proof}
	See \cite[Claim 2]{SCC.Loria.Panteley.ea2005}
\end{proof}

The Matrosov theorem can now be stated as follows.
\begin{thrm}\label{thm:Matrosov}
	Let $0\in\mathcal{D}$ and let $F:\Omega\rightrightarrows\R^{n}$ be a set-valued map with compact values such that (\ref{eq:Inclusion}) admits solutions over $ \Omega $ and is uniformly stable at $x=0$. If, for each pair of numbers $\delta,\Delta\in\R$, such that $0\leq\delta\leq\Delta$ and $\mathrm{D}\left(\delta,\Delta\right)\subset\mathcal{D}$, there exist $ \mathcal{U} - $reduced Matrosov functions for $\left(F,\delta,\Delta\right)$, then (\ref{eq:Inclusion}) is uniformly asymptotically stable at $x=0$. If $\mathcal{D}=\R^{n}$ and if (\ref{eq:Inclusion}) is uniformly globally stable at $x=0$ then (\ref{eq:Inclusion}) is uniformly globally asymptotically stable at $x=0$.
\end{thrm}
\begin{proof}
	Select $\Delta>0$ such that $\overline{\B}\left(0,\Delta\right)\subset\mathcal{D}$ and let $r>0$ be such that
	\begin{equation}
		x\left(\cdot\right)\in \mathscr{S}\left(\overline{\B}\left(0,r\right)\times\R_{\geq 0}\right)\implies x\left(t\right)\in\overline{\B}\left(0,\Delta\right), \forall t\in\R_{\geq t_{0}}. \label{eq:rSelect}
	\end{equation}
	Let $\epsilon\in\left(0,r\right)$ and select $\delta>0$ such that
	\begin{equation}
		x\left(\cdot\right)\in \mathscr{S}\left(\overline{\B}\left(0,\delta\right)\times\R_{\geq 0}\right) \implies x\left(t\right)\in\overline{\B}\left(0,\epsilon\right), \forall t\in\R_{\geq t_{0}}. \label{eq:deltaSelect}
	\end{equation}
	By repeated application of Lemmas \ref{lem:Weird 1} and \ref{lem:Weird 2} it can be shown that $\forall\delta>0$, $\exists\zeta>0$ and $K_{1},\cdots,K_{M-1}>0$ such that $\forall\left(z,x\right)\in\overline{\B}\left(0_{m},\gamma\right)\times\mathrm{D}\left(\delta,\Delta\right)$, 
	\begin{equation}
		Z\left(z,x\right)\coloneqq\sum_{j=1}^{M-1}K_{j}Y_{j}\left(z,x\right)+Y_{M}\left(z,x\right)\leq-\frac{\zeta}{2^{M-1}}.\label{eq:Zbound}
	\end{equation}
	Let $W\in\lip\left(\Omega,\R\right)$ be defined as $
	W\left(x,t\right)\coloneqq\sum_{j=1}^{M-1}K_{j}W_{j}\left(x,t\right)+W_{M}\left(x,t\right).$ From Def. \ref{enu:boundedness},
	\begin{equation}
		\left|V\left(x,t\right)\right|\leq\gamma\left(1+\sum_{j=1}^{M-1}K_{j}\right)\eqqcolon\eta.\label{eq:Vbound-2}
	\end{equation}
	
	Fix $\left(x_{0},t_{0}\right)\in\overline{\B}\left(0,r\right)\times\R_{\geq 0}$ and $x\left(\cdot\right)\in \mathscr{S}\left(\left\{\left( x_{0},t_{0}\right)\right\} \right)$. The selection of $ r $ in \eqref{eq:rSelect} implies that the solution $ x\left(\cdot\right) $ satisfies $x\left(t\right)\in\overline{\B}\left(0,\Delta\right)$, $\forall t\in\R_{\geq t_{0}}$. From Def. \ref{enu:derivative Bound}, $\dot{\overline{V}}_{\mathcal{U}_{j}}\left(x,t\right)\leq Z\left(\phi\left(x,t\right),x\right)$, $\forall\left(x,t\right)\in\mathrm{D}\left(\delta,\Delta\right)\times\R_{\geq 0}$, and hence, from Thm. \ref{thm:Wbound}, 
	\begin{equation}
		\dot{V}\left(x\left(t\right),t\right)\leq Z\left(\phi\left(x\left(t\right),t\right),x\left(t\right)\right),\label{eq:WdotLessThanZ}
	\end{equation}
	for almost all $t\in x^{-1}\left(\mathrm{D}\left(\delta,\Delta\right)\right)$. Using Def. \ref{enu:boundedness} and (\ref{eq:Zbound}), 
	\begin{equation}
		Z\left(\phi\left(x\left(t\right),t\right),x\left(t\right)\right)\leq-\frac{\zeta}{2^{M-1}},\label{eq:Znegative}
	\end{equation}
	for almost all $t\in x^{-1}\left(\mathrm{D}\left(\delta,\Delta\right)\right)$.
	
	Let $T>\frac{2^{M}\eta}{\zeta}$. The claim is that $\left\Vert x\left(t\right)\right\Vert \leq\epsilon$, $\forall t\in\R_{\geq t_{0}+T}$. If not, then the selection of $ \delta $ in \eqref{eq:deltaSelect} implies that $x\left(t\right)\in\mathrm{D}\left(\delta,\Delta\right)$, $\forall t\in\left[t_{0},t_{0}+T\right]$. Hence, from (\ref{eq:WdotLessThanZ}) and (\ref{eq:Znegative}),
	\begin{equation}
		\dot{V}\left(x\left(t\right),t\right)\leq-\frac{\zeta}{2^{M-1}},\label{eq:WDotNegative}
	\end{equation}
	for almost all $t\in\left[t_{0},t_{0}+T\right]$. Integrating (\ref{eq:WDotNegative}) and using the bound in (\ref{eq:Vbound-2}), $\frac{T\zeta}{2^{M-1}}\leq2\eta$, which contradicts $T>\frac{2^{M}\eta}{\zeta}$. Hence, $\forall\epsilon\in\left(0,r\right)$, $\exists T>0$ such that $x\left(\cdot\right)\in \mathscr{S}\left(\overline{\B}\left(0,r\right)\times\R_{\geq 0}\right)$ $\implies$ $\left\Vert x\left(t\right)\right\Vert <\epsilon$, $\forall t\in\R_{\geq t_{0}+T}$, i.e., (\ref{eq:Inclusion}) is uniformly asymptotically stable at $x=0$.
	
	If $\mathcal{D}=\R^{n}$ and if (\ref{eq:Inclusion}) is uniformly globally stable at $x=0$ then $r$ can be selected arbitrarily large, and hence, the result is global.
\end{proof}
The following example demonstrates an application of the Matrosov theorem.
\begin{xmpl}
\label{ex:Matrosov}Let $H:\R\rightrightarrows\R$ be defined as in Example \ref{exa:Nonaut} and let $F:\R^{2}\times\R_{\geq 0}\rightrightarrows\R^{2}$ be defined as in Example \ref{exa:KLYT}. Let $U_{1}$ be defined as in (\ref{eq:U}). Let $W_{1}:\R^{2}\times\R_{\geq 0}\to\R$ be defined as $W_{1}\left(x,t\right)\coloneqq x_{1}^{2}+\left(1+g\left(t\right)\right)x_{2}^{2}$. It follows that $\dot{\overline{W}_{1}}_{\left\{ U_{1}\right\} }\left(x,t\right)\leq-2x_{2}^{2}$, $\forall \left(x,t\right)\in\R^{2}\times\R_{\geq 0}$, and uniform global stability of (\ref{eq:Inclusion}) at $x=0$ can be concluded from Thm. \ref{thm:Nonautonomous}.

Let $\phi\left(x,t\right)=0$, $\forall\left(x,t\right)\in\R^{2}\times\R_{\geq 0}$ and let $Y_{1}\left(z,x\right)\coloneqq-2x_{2}^{2}$, $\forall \left(z,x\right)\in\R\times\R^{2}$. Let $W_{2}\left(x,t\right)\coloneqq x_{1}x_{2}$. The function $U_{2}:\R^{2}\times\R_{\geq 0}\to\R$, defined as (see Fig. \ref{fig:U2})
\begin{figure}
\begin{center}
	\includegraphics[width=0.5\columnwidth]{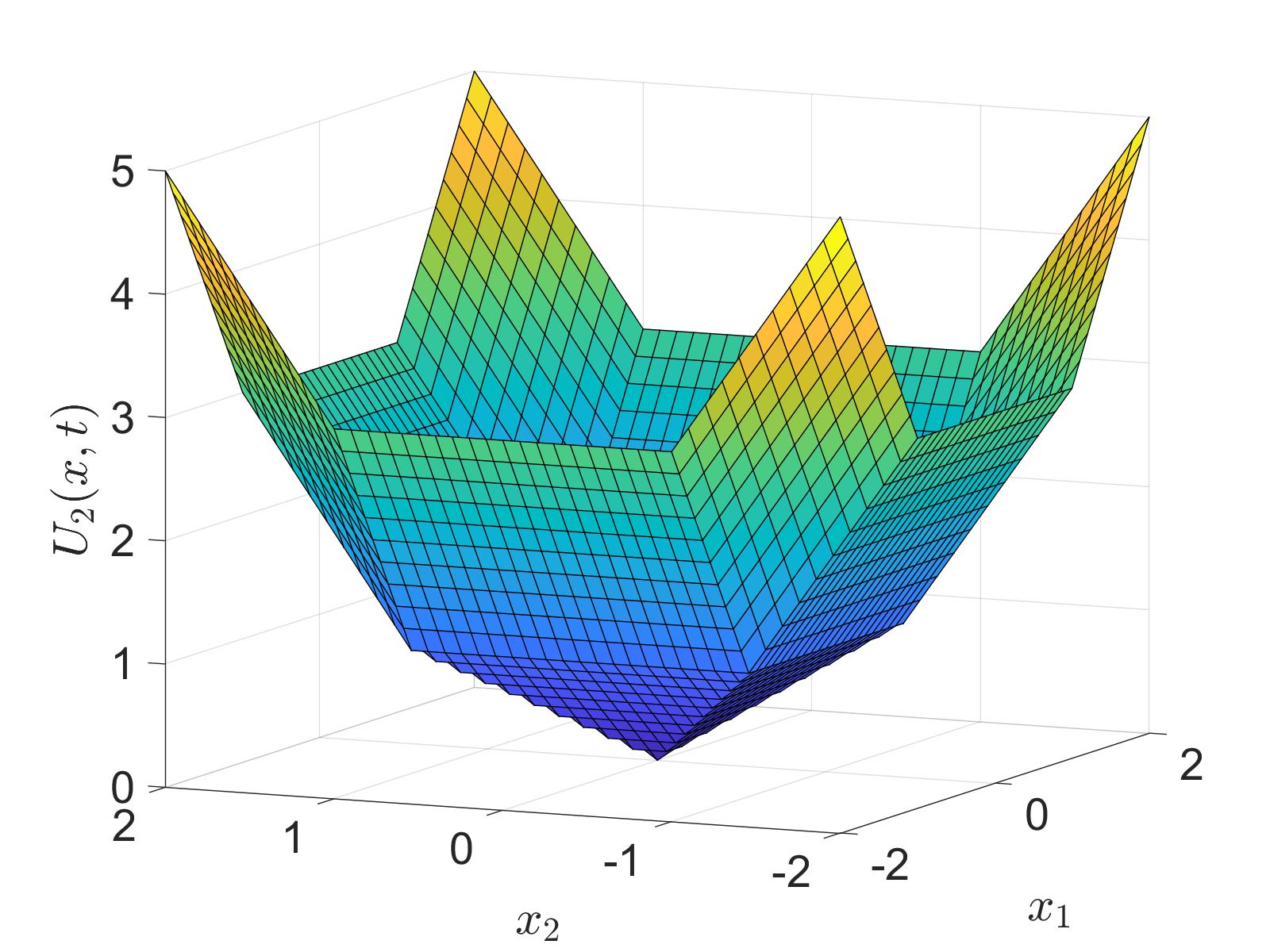}
	\caption{\label{fig:U2}A snapshot of the function $U_{2}:\R^{2}\times\R_{\geq 0}\to\R$.}
\end{center}
\end{figure}
\[
	U_{2}\left(x,t\right)=\begin{cases}
		\left|x_{1}\right| & \left|x_{1}\right|>\left|x_{2}\right|\land x\in\mathrm{Sq},\\
		\left|x_{2}\right| & \left|x_{1}\right|\leq\left|x_{2}\right|\land x\in\mathrm{Sq},\\
		1+U^{*}\left(x,t\right) & x\notin\mathrm{Sq},
	\end{cases}
\]
where
\[
	U^{*}\left(x,t\right)=\max\left(\left(2x_{1}-2\right),0\right)-\min\left(\left(2x_{1}+2\right),0\right)+\max\left(\left(2x_{2}-2\right),0\right)-\min\left(\left(2x_{2}+2\right),0\right), \forall \left(x,t\right)\in\R^{2}\times\R_{\geq 0}
\]
and `$\mathrm{Sq}$' denotes the open unit square centered at the origin, satisfies $U_{2}\in\lip\left(\R^{2}\times\R_{\geq 0},\R\right)$. In addition, since $ U_{2} $ is convex, it is also regular \cite[Prop. 2.3.6]{SCC.Clarke1990}. The Clarke gradient of $U_{2}$ is given by
\[
	\partial U_{2} \left(x,t\right) = \begin{cases}
		\left[\left\{2\sgn1\left(x_{1}\right)\right\} \,\, ; \,\, \left\{2\sgn1\left(x_{2}\right)\right\} \,\, ; \,\, \left\{ 0\right\}\right] & \left|x_{1}\right|\neq1\land\left|x_{2}\right|\neq1,\\
		\left[\overline{\co}\left\{0,2\sgn\left(x_{1}\right)\right\} \,\, ; \,\, \left\{2\sgn1\left(x_{2}\right)\right\} \,\, ; \,\, \left\{ 0\right\}\right] & \left|x_{1}\right|=1\land\left|x_{2}\right|\neq1,\\
		\left[\left\{2\sgn1\left(x_{1}\right)\right\} \,\, ; \,\, \overline{\co}\left\{0,2\sgn\left(x_{2}\right)\right\} \,\, ; \,\, \left\{ 0\right\}\right] & \left|x_{1}\right|\neq1\land\left|x_{2}\right|=1,
	\end{cases}
\]
if $x\notin\overline{\mathrm{Sq}}$,
\[
	\partial U_{2} \left(x,t\right) = \begin{cases}
		\left[\left[-1,1\right] \,\, ; \,\, \left[-1,1\right] \,\, ; \,\, \left\{ 0\right\}\right] & \left|x_{1}\right|=0\land\left|x_{2}\right|=0,\\
		\left[\sgn\left(x_{1}\right) \,\, ; \,\, \left\{ 0\right\} \,\, ; \,\, \left\{ 0\right\}\right]  & \left|x_{1}\right|>\left|x_{2}\right|,\\
		\left[\left\{ 0\right\} \,\, ; \,\, \sgn\left(x_{2}\right) \,\, ; \,\, \left\{ 0\right\}\right] & \left|x_{1}\right|<\left|x_{2}\right|,\\
		\left[\overline{\co}\left\{ 0,\sgn\left(x_{1}\right)\right\} \,\, ; \,\, \overline{\co}\left\{0,\sgn\left(x_{2}\right)\right\} \,\, ; \,\, \left\{ 0\right\}\right] & \left|x_{1}\right|=\left|x_{2}\right|>0,
	\end{cases}
\]
if $x\in\mathrm{Sq}$, and 
\[
	\partial U_{2} \left(x,t\right) = \begin{cases}
		\left[\overline{\co}\left\{ \sgn1\left(x_{1}\right), 2\sgn1\left(x_{1}\right)\right\} \,\, ; \,\, 
		\overline{\co}\left\{\sgn1\left(x_{2}\right),2\sgn1\left(x_{2}\right)\right\}  \,\, ; \,\, \left\{ 0\right\}\right] & \left|x_{1}\right|\neq\left|x_{2}\right|,\\
		\left[\overline{\co}\left\{\sgn\left(x_{1}\right),2\sgn\left(x_{1}\right),0\right\} \,\, ; \,\, 
		\overline{\co}\left\{\sgn\left(x_{2}\right),2\sgn\left(x_{2}\right),0\right\} \,\, ; \,\, \left\{ 0\right\}\right] & \left|x_{1}\right|=\left|x_{2}\right|,
	\end{cases}
\]
if $x\in\mathrm{bd}\left(\mathrm{Sq}\right)$. The $\left\{ U_{2}\right\} -$reduced set-valued map corresponding to $F$ is given by
\[
	\tilde{F}_{\left\{ U_{2}\right\} }\left(x,t\right)=\begin{cases}
		F\left(x,t\right) & x\notin\overline{\mathrm{Sq}}\land\left|x_{1}\right|\neq1\land\left|x_{2}\right|\neq1,\\
		F\left(x,t\right) & x\in\mathrm{Sq}\land\left|x_{1}\right|\neq\left|x_{2}\right|,\\
		\left\{ 0\right\}  & x\in\mathrm{Sq}\land\left|x_{1}\right|=0\land\left|x_{2}\right|=0,\\
		\emptyset & \textnormal{otherwise}.
	\end{cases}
\]
The $\left\{ U_{2}\right\} -$generalized derivative of $W_{2}$ is then given by
\[
	\dot{\overline{W}}_{2\left\{ U_{2}\right\} }\left(x,t\right)=\begin{cases}
		\left[x_{2} \quad x_{1} \quad 0\right]\left[x_{2}h\left(t\right) \,\, ; \,\, -x_{1}-x_{2} \,\, ; \,\, 1\right] & \begin{gathered}
		x\notin\overline{\mathrm{Sq}}\land 	\left|x_{1}\right|\neq1\land\left|x_{2}\right|\neq1\\\lor x\in\mathrm{Sq}\land\left|x_{1}\right|\neq\left|x_{2}\right|,\end{gathered}\\\vspace{-2ex}\\
		0 & x\in\mathrm{Sq}\land \left|x_{1}\right|=0\land 	\left|x_{2}\right|=0,\\
		-\infty & \textnormal{otherwise}.
	\end{cases}
\]
That is, $\dot{\overline{W}}_{2\left\{ U_{2}\right\} }\left(x,t\right)\leq-x_{1}^{2}-x_{2}x_{1}+2x_{2}^{2}, \forall \left(x,t\right) \in \R^{2} \times \R_{\geq 0} $. If $Y_{2}\left(z,x\right)\coloneqq-x_{1}^{2}-x_{2}x_{1}+2x_{2}^{2}$, $\forall \left(z,x\right)\in\R\times\R^{2}$, then the functions $\left\{ Y_{1},Y_{2}\right\} $ have the Matrosov property. Furthermore, since $W_{1},W_{2}\in\mathcal{C}^{0}\left(\R^{2}\times\R_{\geq 0},\R\right)$, $\forall0<\delta<\Delta$, $\exists\gamma>0$ such that $\left|W\left(x,t\right)\right|\leq\gamma,$ $\forall\left(x,t\right)\in\mathrm{D}\left(\delta,\Delta\right)\times\R_{\geq 0}$. Hence, $\left\{ W_{1},W_{2}\right\} $ are $ \mathcal{U} - $reduced Matrosov functions for $\left(F,\delta,\Delta\right)$, $\forall0<\delta<\Delta$. Hence, by Thm. \ref{thm:Matrosov}, \eqref{eq:Inclusion} is uniformly globally asymptotically stable at $x=0$.
\end{xmpl}

\section{Conclusion\label{sec:Conclusion}}

This paper demonstrates that locally Lipschitz, regular functions can be used to identify infeasible directions in set-valued maps that define differential inclusions. The infeasible directions can then be removed to yield a point-wise smaller (in the sense of set containment) set-valued map that defines an equivalent differential inclusion. The reduction process results in a novel generalization of the set-valued derivative for locally Lipschitz candidate Lyapunov functions. Statements of Lyapunov stability theorems, invariance theorems, invariance-like results, and Matrosov theorems for differential inclusions that are less conservative than those available in the literature are developed using reduced set-valued maps.

The fact that arbitrary locally Lipschitz, regular functions can be used to restrict differential inclusions to smaller sets of admissible directions indicates that there may be a \emph{smallest} set of admissible directions corresponding to each differential inclusion. Further research is needed to establish the existence of such a set and to find a representation of it that facilitates computation.

\bibliographystyle{elsarticle-num}
\bibliography{sccmaster,scc,scctemp}

\end{document}

%% file: Macros.tex
\newcommand*\diff{\mathop{}\!\mathrm{d}}

\newcommand*{\sgn}{\operatorname{sgn}}

\newcommand*{\R}{\mathbb{R}}

\newcommand*{\co}{\operatorname{co}}

\newcommand*{\B}{\operatorname{B}}

\newcommand*{\N}{\mathbb{N}}

\newcommand*{\e}{\mathrm{e}}

\newcommand*{\dist}{\operatorname{dist}}

\newcommand*{\ropen}[1]{\left[#1\right)}

\newcommand*{\lopen}[1]{\left(#1\right]}

\newcommand*{\lip}{\operatorname{Lip}}

\newcommand*\defnEnd{\leavevmode\unskip\penalty9999 \hbox{}\nobreak\hfill\quad\hbox{$\triangle$}} % works without amsthm

\newcommand*{\pushedTriangle}{% requires amsthm
	\renewcommand\qedsymbol{$\triangle$}
	\pushQED{\qed}
		\qedhere
	\popQED
}